\documentclass[12pt]{article}
\usepackage{amsfonts,amssymb,amsmath,fullpage}

\newcommand{\R}{{\mathbb R}}
\newcommand{\Z}{{\mathbb Z}}

\newcommand{\cB}{{\cal B}}

\newcommand{\Fix}{{\rm Fix}}
\newcommand{\mbe}{\mbox{\boldmath${\eta}$}}

\newcommand{\mbi}{\mbox{\boldmath${\infty}$}}
\newcommand{\mbel}{{\boldmath${\eta}$}}
\newcommand{\mbzl}{{\boldmath${\zeta}$}}
\newcommand{\mbil}{{\boldmath${\infty}$}}
\newcommand{\rea}{\hbox{Re}}
\newcommand{\ima}{\hbox{Im}}

\newtheorem{definition}{Definition}
\newtheorem{theorem}{Theorem}
\newtheorem{lemma}{Lemma}

\newcommand{\proof}{\noindent {\bf Proof: }}

\newcommand{\mi}{\medskip\noindent}
\newcommand{\qed}{\hfill{\bf QED}
\vspace{5mm}}

\begin{document}

\title{Stability and bifurcations of heteroclinic cycles of type Z}

\author{Olga Podvigina\\
UNS, CNRS, Lab. Lagrange, OCA,\\
BP~4229, 06304 Nice Cedex 4, France, and\\
Institute of Earthquake Prediction Theory and\\
Mathematical Geophysics of the Russian Academy of Sciences,\\
84/32 Profsoyuznaya St, 117997 Moscow, Russian Federation
}

\maketitle

\begin{abstract}
Dynamical systems that are invariant under the action of a non-trivial
symmetry group can possess structurally stable heteroclinic cycles.
In this paper we study stability properties of a class of structurally stable
heteroclinic cycles in $\R^n$ which we call {\it heteroclinic
cycles of type Z}. It is well-known that a heteroclinic cycle that is not
asymptotically stable can attract nevertheless a positive measure set from its
neighbourhood. We say that an invariant set $X$ is {\it fragmentarily asymptotically
stable}, if for any $\delta>0$ the measure of its local basin of attraction
${\cal B}_{\delta}(X)$ is positive. A local basin of attraction
${\cal B}_{\delta}(X)$ is the set of such points that trajectories
starting there remain in the $\delta$-neighbourhood of $X$ for all $t>0$, and
are attracted by $X$ as $t\to\infty$. Necessary and sufficient conditions
for fragmentary asymptotic stability are expressed in terms of eigenvalues
and eigenvectors of transition matrices. If all transverse eigenvalues of
linearisations near steady states involved in the cycle are negative, then
fragmentary asymptotic stability implies asymptotic stability. In the latter case
the condition for asymptotic stability is that the transition matrices have an
eigenvalue larger than one in absolute value.
Finally, we discuss bifurcations occurring when the conditions for asymptotic
stability or for fragmentary asymptotic stability are broken.

\noindent
%\bigskip{\bf Key words:} Heteroclinic Cycle, Stability, Symmetry, Bifurcations
\end{abstract}

\section{Introduction}\label{sec_intro}

A smooth dynamical system
\begin{equation}\label{eq_ode}
\dot{\bf x}=f({\bf x}),\quad f:\R^n\to\R^n
\end{equation}
can possess various kinds of invariant sets -- steady states, periodic orbits,
tori, heteroclinic cycles and strange attractors.
Conditions for asymptotic stability and (local) bifurcations of steady
states and periodic orbits are well known (see e.g. \cite{gukhol}). For a
steady state the conditions for
stability are formulated in terms of eigenvalues of the linearisation near
the steady state. For a periodic orbit they are expressed in terms of
eigenvalues of linearisation of the Poincar\'e return map near the periodic
orbit. When the conditions for stability cease to be satisfied, a bifurcation of
the steady state or of the periodic orbit takes place. No complete theory for
stability and bifurcations of heteroclinic cycles is yet available.

Let $\xi_1,\ldots,\xi_m\in\R^n$ be hyperbolic equilibria of (\ref{eq_ode})
and $\kappa_j:\xi_j\to\xi_{j+1}$, $j=1,\ldots,m$,
$\xi_{m+1}=\xi_1$, be a set of trajectories from $\xi_j$ to $\xi_{j+1}$. The
union of the equilibria and the connecting trajectories is called a
{\it heteroclinic cycle}. Generically heteroclinic cycles are structurally
unstable, because an arbitrary small perturbation of $f$
breaks a connection between two saddle steady states. However, the connections
can be structurally stable (or robust) if the dynamical system has a non-trivial
symmetry group and only symmetric perturbations are considered
\cite{am02,Kru97,sot03}, or if the system is constrained to preserve
certain invariant subspaces \cite{Kru97}.

Heteroclinic cycles that are not asymptotically stable can attract
a positive measure set from its small neighbourhood
\cite{ac10,bra94,dh09,ks94,km04,pos10}. We call such heteroclinic cycles
{\it fragmentarily asymptotically stable}. In earlier papers several types of
stability were employed to describe locally attracting, but not asymptotically
%stable invariant sets: {\it essential asymptotic stability} (e.a.s.) (different
%definitions of e.a.s. sets are given by different authors -- cf.
%\cite{Mel91,km95b} and \cite{ac10,bra94}), {\em relative
stable invariant sets: {\it essential asymptotic stability}
\cite{ac10,bra94,km95b,Mel91}, {\em relative
asymptotic stability} \cite{bra94,Ura64}, {\em predominant asymptotic stability}
\cite{pa11}. If a set is stable in any of these senses, then it
is fragmentarily asymptotically stable. If a heteroclinic cycle is not
fragmentarily asymptotically stable, we call it {\it completely unstable}.

Asymptotic stability or fragmentary asymptotic stability of
structurally stable heteroclinic
cycles was considered in a number of papers. A sufficient condition for
asymptotic stability of heteroclinic cycles is given in \cite{km95a}.
A heteroclinic cycle is called {\em simple}, if all eigenvalues of $df(\xi_j)$
are different and the connecting orbits $\kappa_j$ are one-dimensional.
Necessary and sufficient conditions for asymptotic stability of simple
homoclinic and heteroclinic cycles in $\R^4$ are given in \cite{ckms,dh09a,km04};
necessary and sufficient conditions for fragmentary asymptotic stability of simple
heteroclinic cycles in $\R^4$ are given in \cite{pa11} (the term
fragmentary asymptotic stability is not used there). Conditions for asymptotic
stability, essential asymptotic stability or relative asymptotic stability for
heteroclinic cycles is particular systems are presented in
\cite{ac10,bra94,ckms,dh09,dh09a,fs91,ks94,km95b,pos10,pd10}. In some of
these papers \cite{ckms,dh09,dh09a,pos10,pd10} bifurcations of homoclinic
and heteroclinic cycles are also studied.

In the present paper we introduce a class of (structurally stable simple)
heteroclinic cycles in $\R^n$. All simple heteroclinic cycles studied
in the papers cited above, except for the so-called type A cycles, belong
to this class. We call this class {\it type Z heteroclinic cycles}.

For type Z heteroclinic cycles we derive necessary and sufficient conditions
for asymptotic stability and fragmentary asymptotic stability. A cycle is
fragmentarily asymptotically stable, whenever certain inequalities on eigenvalues and
eigenvectors of transition matrices associated with the cycle are
satisfied. If for all $j$ all transverse eigenvalues of $df(\xi_j)$ are
negative and the cycle is fragmentarily asymptotically stable, then it is
asymptotically stable. For each inequality determining asymptotic stability
or fragmentary asymptotic stability we discuss, a bifurcation of which kind
happens when the inequality ceases to be satisfied as a control parameter is
varied.

\section{Definitions}\label{sec_defs}

\subsection{Stability}\label{sec_stab}

Denote by $\Phi_t({\bf x})$ a trajectory of the system (\ref{eq_ode}) starting
at point ${\bf x}$. For a set $X$ and a number $\epsilon>0$, an
$\epsilon$-neighbourhood of $X$ is the set of points satisfying
\begin{equation}\label{ep_nei}
B_{\epsilon}(X)=\{{\bf x}\in R^n:\ d({\bf x},X)<\epsilon\}.
\end{equation}
Let $X$ be a compact invariant set of (\ref{eq_ode}). Denote by
$\cB_{\delta}(X)$ its $\delta$-local basin of attraction defined as
\begin{equation}\label{del_bas}
\cB_{\delta}(X)=\{{\bf x}\in\R^n:\ d(\Phi_t({\bf x}),X)<\delta\hbox{ for any }
t\ge0\hbox{ and }\lim_{t\to\infty}d(\Phi_t({\bf x}),X)=0\}.
\end{equation}

\begin{definition}\label{def1}
A compact invariant set $X$ is called \underline{asymptotically stable}, if for any
$\delta>0$ there exists an $\epsilon>0$ such that
$$B_{\epsilon}(X)\subset\cB_{\delta}(X).$$
\end{definition}

\begin{definition}\label{def2}
We call a compact invariant set $X$ \underline{fragmentarily asymptotically
stable}, if for any $\delta>0$
$$\mu(\cB_{\delta}(X))>0.$$
(Here $\mu$ is the Lebesgue measure of a set in $\R^n$.)
\end{definition}

Evidently, if a set is asymptotically stable, then it is
fragmentarily asymptotically stable.

\begin{definition}\label{def3}
A set $X$ is called \underline{completely unstable}, if there exists
$\delta>0$ such that $\mu(\cB_{\delta}(X))=0$.
\end{definition}

Recall definitions of invariant sets which are not asymptotically stable,
but are attractors in a weaker sense.

\begin{definition}\cite{AshTer00,Mil85}
A compact invariant set $X$ is called a \underline{weak attractor}, if
$\mu(\cB(X))>0$ (here $\cB(X)$ denotes the basin of attraction of $X$).
A compact invariant set $X$ is called a \underline{Milnor attractor}, if it is a weak
attractor and any proper compact invariant subset $Y\subset X$ satisfies
$$
\mu(\cB(X)\setminus\cB(Y))>0.
$$
\end{definition}

As proved in \cite{AshTer00}, any weak attractor contains a Milnor
attractor. Due to the inclusion $\cB_{\delta}(X)\subset\cB(X)$ which
takes place for any $\delta>0$, if a set is fragmentarily asymptotically
stable, then it is a weak attractor. The converse implication is, in general,
wrong; it does not hold when $\cB_{\delta}(X)>0$ for
some $\delta>\delta_0$ and $\cB_{\delta}(X)=0$ for all $\delta<\delta_0$.

\subsection{Heteroclinic cycles}\label{defhet}

In this paper we consider dynamical systems that have a group of symmetries,
which we denote by $\Gamma$. A dynamical system (\ref{eq_ode}) is called
$\Gamma$-equivariant, where $\Gamma\subset{\bf O}(n)$, if
$f:\R^n\to\R^n$ is a $\Gamma$-equivariant vector field, i.e.
$$
f(\gamma {\bf x})=\gamma f({\bf x}),\quad\mbox{ for all }\gamma\in\Gamma.
$$
We assume that the group $\Gamma$ is finite.

Let $\xi_1,\ldots,\xi_m$ be hyperbolic equilibria of (\ref{eq_ode}) with stable
and unstable manifolds $W^s(\xi_j)$ and $W^u(\xi_j)$, respectively, and
$\kappa_j=W^u(\xi_j)\cap W^s(\xi_{j+1})\ne\emptyset$, $j=1,\ldots,m$,
$\xi_{m+1}=\xi_1$, be a set of trajectories from $\xi_j$ to $\xi_{j+1}$.

\begin{definition}\label{def4}
A \underline{heteroclinic cycle} is an invariant set $X\subset\R^n$ comprised of
a set of equilibria $\{\xi_1,\ldots,\xi_m\}$ and a set of
connecting orbits $\{\kappa_1,\ldots,\kappa_m\}$.
\end{definition}

Recall that for a group $\Gamma$ acting
on $\R^n$ the {\it isotropy group} of the point $x\in\R^n$ is the subgroup
$$
\Sigma_x=\{\gamma\in\Gamma\ ~:~\ \gamma x=x\},
$$
and a {\it fixed-point subspace} of a subgroup $\Sigma\subset\Gamma$
is the linear subspace
$$
{\rm Fix}(\Sigma)=\{{\bf x}\in\R^n\ ~: \ \sigma {\bf x}={\bf x}\mbox{ for all } \sigma\in\Sigma\}.
$$

\begin{definition}\label{def5}
A heteroclinic cycle is called \underline{structurally stable (or {\it robust})}, if
for any $j$, $1\le j\le m$, there exists a fixed-point subspace
$P_j={\rm Fix}(\Sigma_j)$, where $\Sigma_j\subset\Gamma$, such that
\begin{itemize}
\item $\xi_{j+1}$ is a sink in $P_j$;
\item $\kappa_j\subset P_j$.
\end{itemize}
\end{definition}

We denote $L_j=P_{j-1}\cap P_j$ and the isotropy subgroup of $L_j$ by
$T_j$; evidently, $\xi_j\in L_j$.

\subsection{Eigenspaces, simple cycles and type Z cycles}
\label{locstr}

For a structurally stable heteroclinic cycle, eigenvalues of
$df(\xi_j)$ can be divided into four classes \cite{km95a,km95b,km04}:
\begin{itemize}
\item Eigenvalues with associated eigenvectors in $L_j$ are called {\it radial}.
\item Eigenvalues with associated
eigenvectors in $P_{j-1}\ominus L_j$ are called {\it contracting}.
\item Eigenvalues
with associated eigenvectors in $P_j\ominus L_j$ are called {\it expanding}.
\item Eigenvalues not belonging to any of the three above classes
are called {\it transverse}.
\end{itemize}

\begin{definition}\label{def6}
(adapted from \cite{km04}). We call a robust heteroclinic cycle
$X\in\R^n\setminus \{0\}$ \underline{simple}, if for any $j$
\begin{itemize}
\item all eigenvalues of $df(\xi_j)$ are distinct;
\item $\dim(P_{j-1}\ominus L_j)=1$.
\end{itemize}
\end{definition}

Note that definitions of {\it simple} heteroclinic cycles different from
the one in \cite{km04} can be found in literature. According to Hofbauer and
Sigmund \cite{hs98}, a heteroclinic cycle is simple, if for each $j$ the
linearisation $df(\xi_j)$ has only one expanding eigenvector and
all transverse eigenvalues are negative. Field \cite{fi96} calls a heteroclinic
cycle simple, if all transverse eigenvalues are negative and
the cycle is a compact set.

\medskip
Denote by $P_j^{\perp}$ the orthogonal complement to $P_j$ in $\R^n$.

\begin{definition}\label{def7}
We call a simple robust heteroclinic cycle $X$ to be of \underline{type Z},
if for any $j$
\begin{itemize}
\item $\dim P_j=\dim P_{j+1}$;
\item the isotropy subgroup of $P_j$, $\Sigma_j$, decomposes
$P_j^{\perp}$ into 1-dimensional isotypic components.
\end{itemize}
\end{definition}

The letter Z in the name of the cycle is chosen as the ``opposite'' one to A:
Consider four eigenspaces that are associated with the dominant (i.e., having
the largest real part) contracting eigenvalue of $df(\xi_j)$; the weakest
(having the smallest real part) transverse eigenvalue of $df(\xi_j)$;
the dominant expanding eigenvalue of $df(\xi_{j+1})$; and the weakest
transverse eigenvalue of $df(\xi_{j+1})$. A heteroclinic cycle is called to be
of type A, if all the four eigenspaces belong to the same $\Sigma_j$-isotypic
component \cite{km95a,pa11}. This isotypic component of $P_j^{\perp}$ is
therefore at least two-dimensional. For type Z cycles, by contrast, all
isotypic components of $P_j^{\perp}$ are required to be one-dimensional.

The condition $\dim(P_{j-1}\ominus L_j)=1$ implies that for any $j$ the
contracting eigenspace at $\xi_j$ is one-dimensional. Together with the condition
$\dim P_j=\dim P_{j+1}$, this implies that the dimension of the expanding
eigenspace is also one. Denote by $n_r$ the number of radial eigenvalues and
by $n_t$ the number of transverse eigenvalues; for a type Z heteroclinic
cycle $n_r$ and $n_t$ are the same for all equilibria, and $n=n_r+n_t+2$.
The radial eigenvalues and the associated eigenvectors near $\xi_j$ are denoted
by $-{\bf r}_j=-\{r_{j,l}\}$ and ${\bf v}_j^r=\{v_{j,l}^r\}$, $1\le l\le n_r$,
the contracting ones by $-c_j$ and $v_j^c$, the expanding ones by $e_j$ and
$v_j^e$, and the transverse ones by ${\bf t}_j=\{t_{j,l}\}$ and
${\bf v}_j^t=\{v_{j,l}^t\}$, $1\le l\le n_t$, respectively.

The basis in $P_j^{\perp}$ can be chosen to be comprised of contracting and
transverse eigenvectors at $\xi_j$, or of expanding and transverse eigenvectors
at $\xi_{j+1}$. For type Z cycles all isotypic components of $P_j^{\perp}$ are
one-dimensional. Since any eigenvector belongs to an isotypic component,
the basis $\{v_{j+1}^e,{\bf v}_{j+1}^t\}$ is a permutation of the basis
$\{v_j^c,{\bf v}_j^t\}$, possibly accompanied by the change of the directions
of some vectors to the opposite ones. Hence, the matrix mapping components
of a vector in the basis $\{v_j^c,{\bf v}_j^t\}$ to components of the vector
in the basis $\{v_{j+1}^e,{\bf v}_{j+1}^t\}$ is a product $A_j^{\pm}A_j$, where
$A_j$ is a permutation matrix, and $A_j^{\pm}$ is a diagonal matrix with
elements $+1$ and $-1$ on the diagonal.

\medskip
In fact, if a dynamical system has a heteroclinic cycle of type Z, then
the system possesses a variety of symmetry-invariant subspaces in addition
to the subspaces $P_j$ required in definition \ref{def6}. Existence
of these subspaces follows from the following lemma.

\begin{lemma}\label{lem0}
Let a group $\Sigma$ act on a linear space $V$. Consider the isotypic
decomposition of the linear space under the action of $\Sigma$:
$$V=U_0\oplus U_1\oplus\ldots\oplus U_K.$$
Suppose
\begin{itemize}
\item the action of $\Sigma$ on $U_0$ is trivial;
\item any $\sigma\in\Sigma$ acts on a $U_k$, $1\le k\le K$, either as $I$ or
as $-I$.
\end{itemize}
Then for any collection of indices $1\le i_1,\ldots,i_l\le K$ there exists
a subgroup $G_{i_1,\ldots,i_l}\subset\Sigma$ such that the subspace
$$V_{i_1,\ldots,i_l}=U_0\oplus U_{i_1}\oplus\ldots\oplus U_{i_l}$$
is a fixed point subspace of the group $G_{i_1,\ldots,i_l}$.
\end{lemma}

\proof
By definition of the isotypic decomposition, the conditions of the lemma
imply that for any $1\le k\le K$ we can find an element $\gamma_k\in\Sigma$ such that
$\gamma_kU_k=-U_k$. For any $0\le k\le K$ and $1\le s\le K$ we can also find
$\gamma_{ks}\in\Sigma$ such that
$$\gamma_{ks}U_k=U_k\hbox{ and }\gamma_{ks}U_s=-U_s.$$
Hence for any $1\le k\le K$ there exists $\sigma_k\in\Sigma$ such that
$$\sigma_kU_k=-U_k\hbox{ and }\sigma_kU_s=U_s\hbox{ for any }s\ne k.$$
(This can be proved by induction in $K$. The proof is omitted.) Evidently,
the subspace $V_{i_1,\ldots,i_l}$ is a fixed point subspace of the subgroup
of $\Sigma$ generated by all $\sigma_k$ with $k\ne i_1,\ldots,i_l$.
\qed

Definition \ref{def6} requires a $\Sigma_j$-invariant map to have no multiple
eigenvalues; consequently, all $\Sigma_j$ satisfy the conditions of the lemma.
Alternatively, by definition of a type Z cycle all isotypic components
of $P_j^{\perp}$ are one-dimensional, hence elements of $\Sigma_j$ act
on these components either as $I$ or $-I$.

\medskip
Following \cite{ks94,km04,pa11}, in order to examine stability
we construct a Poincar\'e map in the vicinity of the cycle.

\subsection{Collection of maps associated with a heteroclinic cycle}
\label{cycmap}

In subsection~\ref{locstr} we have given definitions for radial, contracting,
expanding and transverse eigenvalues of the linearisation $df(\xi_j)$.
Let $(\tilde{\bf u},\tilde v,\tilde w,\tilde{\bf z})$ be local coordinates
near $\xi_j$ in the basis, where radial eigenvectors come the first
(the respective coordinates are $\tilde{\bf u}$), followed by the contracting
and the expanding eigenvectors, the transverse eigenvectors being the last.
Suppose $\tilde\delta$ is small. In a $2\tilde\delta$-neighbourhood of $\xi_j$,
$B_{2\tilde\delta}(\xi_j)$, defined as
$$
B_{2\tilde\delta}(\xi_j)=\{(\tilde{\bf u},\tilde v,\tilde w,\tilde{\bf z})\ :\
\max(|\tilde{\bf u}|,|\tilde v|,|\tilde w|,|\tilde{\bf z}|)<2\tilde\delta\},
$$
system (\ref{eq_ode}) can be approximated by the linear system\footnote{Below
we assume that all eigenvalues are real. Definition \ref{def7} implies that for
type Z cycles transverse, contracting and expanding eigenvalues are real. Radial
eigenvalues can be complex, but this does not change our proof significantly.}
\begin{equation}\label{lmap}
\begin{array}{l}
\dot u_l=-r_{j,l}u_l,\ 1\le l\le n_r\\
\dot v=-c_j v\\
\dot w=e_j w\\
\dot z_s=t_{j,s}z_s,\ 1\le s\le n_t.
\end{array}
\end{equation}
We denote by $({\bf u},v,w,{\bf z})$ the scaled coordinates
$({\bf u},v,w,{\bf z})=
(\tilde {\bf u},\tilde v,\tilde w,\tilde {\bf z})/\tilde\delta$.

Consider a neighbourhood of a steady state $\xi_j$. Let $({\bf u}_0,v_0)$ be
the point in $P_{j-1}$ where trajectory $\kappa_{j-1}$ intersects
with the sphere $|{\bf u}|^2+v^2=1$, and $\bf q$ be local
coordinates in the hyperplane tangent to the sphere at the point
$({\bf u}_0,v_0)$. Coordinates $({\bf u},v)$ of a point in the hyperplane
are related to coordinates $\bf q$ as follows:
\begin{equation}\label{relat}
\left(
\begin{array}{c}
{\bf u}\\
v
\end{array}
\right)={\cal D}^{\|}_j{\bf q}=
\left(
\begin{array}{c}
{\bf u}_0\\
v_0
\end{array}
\right)+D^{\|}_j{\bf q},
\end{equation}
where $D^{\|}_j$ is a $n_r\times(n_r+1)$ matrix. Some components of
${\bf u}_0$ can vanish, if e.g. $\kappa_{j-1}$ belongs to an invariant
subspace in $P_{j-1}$. (Note that $P_j$ is not required to be the smallest
possible subspace.) $v_0$ does not vanish, because it is the component in the
contracting direction.

Near $\xi_j$ we define two crossections of the heteroclinic cycle. One,
denoted by $\tilde H^{(in)}_j$, is an $(n-1)$-dimensional hyperplane
intersecting connection $\kappa_{j-1}$ at the point $({\bf u}_0,v_0,0,0)$;
coordinates in the hyperplane are $({\bf q},w,{\bf z})$. Another one,
$\tilde H^{(out)}_j$, is parallel to the hyperplane $w=0$ and intersects
connection $\kappa_j$ at the point $w=1$; coordinates in the hyperplane are
$({\bf u},v,{\bf z})$. Near $\xi_j$ trajectories of the system (\ref{eq_ode})
can be approximated by a local map (called the {\em first return
map})\footnote{If some components $t_{j,s}$ of ${\bf t}_j$ are positive, then
the local map is defined for $z_s$ satisfying the inequality
$|z_s|<K(1-\delta)|w|^{t_{j,s}/e_j}$, where $K$ is a constant and $\delta$ is
small (see \cite{ks94,pa11}). However, this restriction is not important,
because in order to study stability of a cycle we study stability of a fixed
point of a collection of maps $\R^N\to\R^N$, and the maps are defined for all
$x\in\R^N$. Moreover, the local map is defined only for particular
signs of $v$ and $w$. To overcome this complication, we consider group orbits
of heteroclinic cycles, see subsection \ref{defcyc}.}
$\phi_j:\tilde H^{(in)}_j\to \tilde H^{(out)}_j$ relating a point, where a
trajectory enters the neighbourhood, to the point, where it exits.
In the leading order (see (\ref{lmap}) and (\ref{relat})\,), the local map is
\begin{equation}\label{phmap}
\phi_j(\{q_l\},w,\{z_s\})=(\{(u_l+\sum_{s=1}^{n_r}
D_{j,ls}^{\|}q_s)w^{r_{j,l}/e_j}\},v_0w^{c_j/e_j},\{z_sw^{-t_{j,s}/e_j}\}).
\end{equation}
The map can be expressed as a superposition
$\phi_j={\cal C}^{\rm tot}_j{\cal D}^{\rm tot}_j$, where
${\cal C}^{\rm tot}_j:R^n\to R^{n-1}$ and ${\cal D}^{\rm tot}_j:R^{n-1}\to R^n$.
The action of the map ${\cal D}^{\rm tot}_j$ on the $\bf q$ coordinates is
presented by (\ref{relat}), and its action on the $w$ and $z$ coordinates is
trivial. The map ${\cal C}^{\rm tot}_j$ is
\begin{equation}\label{mapc}
{\cal C}^{\rm tot}_j
\left(
\begin{array}{c}
\{u_l\}\\
v\\
w\\
\{z_s\}
\end{array}
\right)
=
\left(
\begin{array}{c}
\{u_lw^{r_{j,l}/e_j}\}\\
vw^{c_j/e_j}\\
\{ z_sw^{-t_{j,s}/e_j}\}
\end{array}
\right).
\end{equation}

Near connection $\kappa_j$ the system (\ref{eq_ode}) can be approximated
by a global map (also called a {\em connecting diffeomorphism})
$\psi_j:\tilde H^{(out)}_j\to \tilde H^{(in)}_{j+1}$,
\begin{equation}\label{cdif}
\left(
\begin{array}{c}
{\bf q}^{j+1}\\
w^{j+1}\\
{\bf z}^{j+1}
\end{array}
\right):=\psi_j
\left(
\begin{array}{c}
{\bf u}^j\\
v^j\\
{\bf z}^j
\end{array}
\right)=A^{\rm tot}_jB^{\rm tot}_j
\left(
\begin{array}{c}
{\bf u}^j\\
v^j\\
{\bf z}^j
\end{array}
\right),
\end{equation}
where superscripts in the notation of components indicate, whether
the respective vector is decomposed
in the local basis near $\xi_j$ or near $\xi_{j+1}$. The $(n-1)\times(n-1)$
matrix $B^{\rm tot}_j$ presents the map $\psi_j$ in the local coordinates near
$\xi_j$ (i.e., the basis near $\xi_{j+1}$ is the same, as near $\xi_j$,
and the origin is shifted to $\xi_{j+1}$), and matrix $A^{\rm tot}_j$
relates the coordinates in the two local bases). Each matrix is comprised
of two diagonal blocks (the respective non-diagonal blocks vanish).
The first $n_r\times n_r$ blocks, $B^{\|}_j$ and $A^{\|}_j$, approximate the
maps acting in $P_j$, and the second blocks, $B_j$ and $A^{\perp}_j$,
the maps acting in $P_j^{\perp}$.
Lemma \ref{lem0} implies that the matrix $B_j$ is diagonal.
As discussed in subsection \ref{locstr}, $A^{\perp}_j=A_j^{\pm}A_j$.

Denote the superpositions of the local, $\phi_j$, and global, $\psi_j$, maps by
$\tilde g_j=\psi_j\circ\phi_j:\tilde H^{(in)}_j\to\tilde H^{(in)}_{j+1}$.
The Poincar\'e map
$\tilde H^{(in)}_1\to\tilde H^{(in)}_1$ for the cycle is the superposition
$\tilde\pi_1=\tilde g_m\circ\ldots\circ\tilde g_1$; for $j>1$ the Poincar\'e
maps $\tilde H^{(in)}_j\to\tilde H^{(in)}_j$ are constructed similarly:
$$\tilde\pi_j=\tilde g_{j-1}\circ\ldots\circ\tilde g_1\circ\tilde g_m
\circ\ldots\circ\tilde g_j.$$

The coordinates $(w,{\bf z})$ used to define the maps $\tilde g_j$ are
independent of $\bf q$. Hence, we can define maps $g_j$ which are
restrictions of the maps $\tilde g_j$ into the $(w,{\bf z})$-subspace:
\begin{equation}\label{eq_mapg0a}
g_j(w,{\bf z})=
A_j^{\pm}A_jB_j
\left(
\begin{array}{c}
v_0w^{c_j/e_j}\\
\{z_sw^{-t_{j,s}/e_j}\}
\end{array}
\right).
\end{equation}
We call the set of maps $\{g_1^m\}=\{g_1,\ldots,g_m\}$, where
$g_j:\R^{n_t+1}\to\R^{n_t+1}$ have been constructed above,
{\it a collection of maps associated with the heteroclinic cycle}
$\{\xi_1,\ldots,\xi_m\}$. The collection of maps
$\{g_l^{l-1}\}=\{g_l,\ldots,g_m,g_1,\ldots,g_{l-1}\}$ is associated with
the heteroclinic cycle $\{\xi_l,\ldots,\xi_m,\xi_1,\ldots,\xi_{l-1}\}$
which geometrically coincides with the former cycle.

\subsection{Comments on definitions of a heteroclinic cycle}
\label{defcyc}

Note that a heteroclinic cycle satisfying the commonly used definition
\ref{def4} is never asymptotically stable in the sense of definition \ref{def1}
for the following reasons. Denote by $\gamma^{(out)}$ a symmetry which
belongs to $T_j$ but does not belong to $\Sigma_{j-1}$. The symmetry fixes
$\xi_j$, reverses the
sign of $w$, maps $\tilde H^{(in)}_j$ into itself, and maps the connection
$\kappa_j$ into a different connection $\gamma^{(out)}\kappa_j$ exiting a
neighbourhood of $\xi_j$ via $\gamma^{(out)}\tilde H^{(out)}_j$ (which differs
from $\tilde H^{(out)}_j$ by the sign of $w$). Hence, for any small $\epsilon$
we can find points in $B_{\epsilon}(X)\cap\tilde H^{(in)}_j$ such that
trajectories starting there do not follow the connection $\kappa_j$.
Similarly, the sign of $v$ is the same throughout $\tilde H^{(in)}_j$,
and the symmetry $\gamma^{(in)}\in T_j$ such that
$\gamma^{(in)}\not\in\Sigma_j$ maps $\tilde H^{(in)}_j$ into a different
crossection with the opposite sign of $v$ and the connection $\kappa_{j-1}$
into $\gamma^{(in)}\kappa_{j-1}$. Hence, the map
$\phi_j:\tilde H^{(in)}_j\to\tilde H^{(out)}_j$ is defined for points
with certain fixed sign of $w$, and its image contains points with
a particular fixed sign of $v$.

For this reason, a heteroclinic cycle is often defined as (a connected
component of) an orbit under the action of the group of symmetries $\Gamma$
of a heteroclinic cycle satisfying definition \ref{def4}.
For a group orbit, the maps $\phi_j$ are defined for all signs of $w$ and
$v$, because in addition to $\kappa_{j-1}$ and $\kappa_j$ the orbit involves
$\gamma^{(in)}\kappa_{j-1}$ and $\gamma^{(out)}\kappa_j$ as well.
A group orbit can be asymptotically stable according to definition \ref{def1}.
Below, when speaking about a heteroclinic cycle we always assume a group orbit.

Consider a sample trajectory close to a heteroclinic cycle. After the trajectory
passes near $\xi_j$, it can head for $\xi_{j+1}$ following connection
$\kappa_j$, or for $\gamma^{(out)}\xi_{j+1}$ following connection
$\gamma^{(out)}\kappa_j$. The choice depends on the sign of $w$. Hence,
different trajectories can visit different sets of equilibria.
Suppose that all subspaces $P_j$ are maximal, i.e. there does not exist
an invariant proper subspace of $P_j$, $\tilde P_j$, such that
$\kappa_j\subset\tilde P_j$. Then generically none of the $({\bf u}_0,v_0)$
components vanish. The signs of components of $\bf z$ are preserved by
the local maps. Any global map preserves the sign of each component of $\bf z$
for all trajectories, or reverses it for all trajectories. Hence, a particular
path along the cycle followed by the trajectory (i.e., the sequence of equilibria
visited by the trajectory) $\Phi_t({\bf x})$ for ${\bf x}\in\tilde H^{(in)}_j$
is uniquely determined by the signs of coordinates of $\bf x$ (i.e., all points
in each orthant of $\R^{n-1}$ of $\tilde H^{(in)}_j$ follow the same path).

When definition \ref{def4} of a heteroclinic cycle is used, often a different
object is tacitly assumed. Suppose there exists a symmetry $\gamma\in\Gamma$
such that the heteroclinic cycle $\{\xi_1,\ldots,\xi_m\}$ can be generated from
its smaller subcycle $\{\xi_1,\ldots,\xi_l\}$, $m=lK$, by applying the symmetry
$\gamma$:
$$\gamma\xi_{s+(k-1)l}=\xi_{s+kl}\hbox{ for all }1\le s\le l\hbox{ and }
1\le k\le K.$$
Then the subcycle $\{\xi_1,\ldots,\xi_l\}$ is called sometimes a heteroclinic
cycle referring to its group orbit.
If $l=1$, then the heteroclinic cycle is called a {\it homoclinic
cycle} and the symmetry $\gamma$ is called a {\it twist} \cite{am02,sot03}.

\subsection{Collection of maps: definitions of stability}
\label{stamap}

For a collection of maps $\{g_1^m\}=\{g_1,\ldots,g_m\}$ we define superpositions
\begin{equation}\label{mapg}
\pi_j=g_{j-1}\circ\ldots\circ g_1\circ g_m\circ\ldots\circ g_{j+1}\circ g_j
\end{equation}
(for $j=1$ and $j=2$ this reduces to
$\pi_1=g_m\circ\ldots\circ g_2\circ g_1$ and
$\pi_2=g_1\circ g_m\circ\ldots\circ g_2$, respectively) and
\begin{equation}\label{sup2}
g_{(l,j)}=\left\{
\renewcommand{\arraystretch}{2.0}
\begin{array}{ll}
g_{l}\circ\ldots\circ g_j, & l>j\\
g_{l}\circ\ldots\circ g_m\circ g_1\circ\ldots\circ g_j, & l<j
\end{array}\right..
\end{equation}

Given a collection of maps $\{g_1^m\}=\{g_1,\ldots,g_m\}$, $g_j:\R^N\to\R^N$,
we define a discrete dynamical system by the relation
${\bf y}_{n+1}=\pi_1{\bf y}_n$. We call ${\bf y}^1\in\R^N$ a fixed point of
the collection of maps $\{g_1^m\}$, if $\pi_1{\bf y}^1={\bf y}^1$. Evidently,
${\bf y}^l=g_{(l-1,1)}{\bf y}^1$ is then
a fixed point of the collection of maps $\{g_l^{l-1}\}$.

\begin{definition}\label{def8}
We say that a fixed point ${\bf y}^1\in\R^N$ of a collection of maps
\begin{equation}\label{colm}
\{g_1^m\}=\{g_1,\ldots,g_m\},\ g_j:\R^N\to\R^N,
\end{equation}
is \underline{asymptotically stable}, if for any $\delta>0$ there exists an
$\epsilon>0$ such that for any $1\le l\le m$
$$d({\bf x},{\bf y}^l)<\epsilon,\hbox{ where } {\bf y}^l=g_{(l-1,1)}{\bf y}^1,$$
implies
%\begin{equation}\label{asg1}
$$d(\pi_j^kg_{(j-1,l)}{\bf x},g_{(j-1,l)}{\bf y}^l)<\delta
\mbox{ for all }\ 1\le j\le m,\ k\ge0$$
%\end{equation}
and
%\begin{equation}\label{asg2}
$$\lim_{k\to\infty}d(\pi_j^kg_{(j-1,l)}{\bf x},g_{(j-1,l)}{\bf y}^l)=0
\mbox{ for all }1\le j\le m.$$
%\end{equation}
\end{definition}

\begin{definition}\label{def9}
We say that a fixed point ${\bf y}^1\in\R^N$ of a collection of maps $\{g_1^m\}$
is \underline{fragmentarily}\break\underline{asymptotically stable}, if
for any $\delta>0$
$$\mu(\cB_{\delta}(\{g_1^m\},{\bf y}^1))>0,$$
where
$$\cB_{\delta}(\{g_1^m\},{\bf y}^1):=
\{{\bf x}~:~{\bf x}\in\R^N,\ d(\pi_j^kg_{(j-1,1)}{\bf x},
g_{(j-1,1)}{\bf y}^1)<\delta\mbox{ for all }1\le j\le m,\ k\ge0$$
$$\hbox{and }\lim_{k\to\infty}d(\pi_j^kg_{(j-1,1)}{\bf x},g_{(j-1,1)}{\bf y}^1)=0
\mbox{ for all }1\le j\le m\}.$$
\end{definition}

\begin{definition}\label{def91}
We say that a fixed point ${\bf y}^1\in\R^N$ of a collection of maps $\{g\}_1^m$ is
\underline{completely unstable}, if there exists $\delta>0$ such that
$$\mu(\cB_{\delta}(\{g_1^m\},{\bf y}^1))=0.$$
\end{definition}

\section{Stability of a cycle and a collection of maps}
\label{stabcm}

In this section we prove two theorems relating the asymptotic stability
of a type Z heteroclinic cycle with the asymptotic stability of the
fixed point $(w,{\bf z})=\bf 0$ of the collection of maps associated
with the cycle. (The point $(w,{\bf z})=\bf 0$ is evidently a fixed point
of the collection of maps constructed in subsection \ref{cycmap}.)

\begin{theorem}\label{th1}
Let $\{g_1^m\}$, $g_j:\R^{n_t+1}\to\R^{n_t+1}$, be the collection of maps
associated with a heteroclinic cycle of type Z. The cycle is
asymptotically stable, if and only if the fixed point $(w,{\bf z})=\bf 0$
of the collection of maps is asymptotically stable.
\end{theorem}

\proof
A necessary condition for asymptotic stability of both the cycle and the
collection of maps is that all transverse eigenvalues are negative. We assume
henceforth in this proof that this is the case.
% Thus, the maps $g_j$ are
%defined for all points in a small neighbourhood of zero in $H_j^{(in)}$.

For a trajectory $\Phi_t({\bf x})$ belonging to $B_{\delta}(X)$ for all $t>0$,
denote by $\Phi_{j,k}^{(in)}({\bf x})$ the $k$-th intersection of
$\Phi_t({\bf x})$ with
$\tilde H_j^{(in)}$, by $\Phi_{j,k}^{(out)}({\bf x})$ the $k$-th intersection of
$\Phi_t({\bf x})$ with $\tilde H_j^{(out)}$, and by $t_{j,k}^{(in)}$ and
$t_{j,k}^{(out)}$ the times of occurrence of the intersections.

For a sufficiently small $\delta>0$ the collection of maps $\{\tilde g_1^m\}$,
where $\tilde g_j:\R^{n-1}\to\R^{n-1}$, accurately approximates trajectories
in the vicinity of the cycle; i.e., if ${\bf x}\in\tilde H_l^{(in)}$ and
$\Phi_t({\bf x})\in B_{\delta}(X)$ for all $t>0$, then
\begin{equation}\label{apmap}
\Phi_{j,k}^{(in)}({\bf x})\approx\tilde \pi_j^{k-1}\tilde g_{j-1,l}{\bf x}.
\end{equation}
Stability of the cycle implies stability of the fixed point $\bf 0$ of the
collection of maps, because\\
$\bullet$ if $\Phi_t({\bf x})\in B_{\delta}(X)$ for any ${\bf x}$ satisfying
$d({\bf x},X)<\epsilon$, then
$$d(\Phi_{j,k}^{(in)}({\bf x}),X)<\delta\hbox{ for any }j,\ k \hbox{ and }
{\bf x}\in \tilde H^{(in)}_l,\ |{\bf x}|<\epsilon,$$
and hence, by (\ref{apmap}),
$|\pi_j^k g_{j-1,l}{\bf x}|<|\tilde \pi_j^k\tilde g_{j-1,l}{\bf x}|<\delta$;\\
$\bullet$ if $\lim_{t\to\infty}d(\Phi_t({\bf x}),X)=0$ for any ${\bf x}$ satisfying
$d({\bf x},X)<\epsilon$, then
$$\lim_{k\to\infty}d(\Phi_{j,k}^{(in)}({\bf x}),X)=0\hbox{ for any }j\hbox{ and }
{\bf x}\in \tilde H^{(in)}_l,\ |{\bf x}|<\epsilon,$$
and hence, by (\ref{apmap}),
$$\lim_{k\to\infty}|\pi_j^k g_{j-1,l}{\bf x}|
<\lim_{k\to\infty}|\tilde \pi_j^k\tilde g_{j-1,l}{\bf x}|=0.$$
Therefore, by definition \ref{def8}, $\bf 0$ is an asymptotically
stable fixed point of the collection of maps.

\medskip
To prove that asymptotic stability of the cycle follows from asymptotic
stability of the fixed point $\bf 0$ of the collection of maps, we first show
that asymptotic stability of the fixed point $(w,{\bf z})=\bf 0$
of the collection $\{g_1^m\}$ implies asymptotic stability of the fixed point
$({\bf q},w,{\bf z})=\bf 0$ of the collection $\{\tilde g_1^m\}$.
Denote by $\tilde g_j^q$ the map $\tilde g_j$ restricted to the subspace
$(w=0,\ {\bf z}=\bf 0)$ which is equipped with the $\bf q$ coordinates.
By virtue of (\ref{relat})-(\ref{mapc}) and (\ref{cdif}), the map takes the form
\begin{equation}\label{mapq}
g_j^q({\bf q},w,{\bf z})=A^{\|}_jB^{\|}_j{\cal C}^{\|}_j{\cal D}^{\rm tot}_j
\left(
\begin{array}{c}
{\bf q}\\
w\\
{\bf z}
\end{array}
\right),
\end{equation}
where
\begin{equation}\label{mape}
{\cal C}^{\|}_j
\left(
\begin{array}{c}
{\bf u}\\
v\\
w\\
{\bf z}
\end{array}
\right)=
(\{u_lw^{r_{j,l}/e_j}\}).
\end{equation}
Let
$K_A^j=n_r\max_{l,s}|A_{j,ls}^{\|}|$, $K_B^j=n_r\max_{l,s}|B_{j,ls}^{\|}|$
$K_D^j=n_r\max(|{\bf u}_0|,1,\max_{l,s}|D_{j,ls}^{\|}|)$ and
$r_{\rm min}^j=\min_l(r_{j,l}/e_j)$. For a given $\delta>0$, choose
$\delta_j$ satisfying
\begin{equation}\label{est2}
K_A^jK_B^jK_D^j\delta_j^{r_{\rm min}^j}<\delta/2.
\end{equation}
Since $\bf 0$ is a stable fixed point of the collection $\{g_1^m\}$,
we can find $\epsilon>0$ such that
\begin{equation}\label{est1}
|\pi_j^kg_{j-1,l}{\bf x}|<\min(\delta/2,\min_s\delta_s)
\mbox{ for all }1\le j,l\le m,\ k\ge0,\ |{\bf x}|<\epsilon.
\end{equation}
If (\ref{est1}) holds true, then
\begin{equation}\label{est4}
|\tilde \pi_j^k\tilde g_{j-1,l}{\bf x}|<\delta\mbox{ for all }
1\le j,l\le m,\ k\ge0,\ |{\bf x}|<\epsilon
\end{equation}
by virtue of (\ref{relat})-(\ref{cdif}) and (\ref{mapq})-(\ref{est2}).
The proof that $\lim_{k\to\infty}\pi_j^kg_{(j-1,l)}{\bf x}=0$ implies\\
$\lim_{k\to\infty}\tilde\pi_j^k\tilde g_{j-1,l}{\bf x}=0$ is similar and
we omit it.

\medskip
Second, we prove that if the fixed point $\bf 0$ of the collection
$\{\tilde g_1^m\}$ is asymptotically stable, then for any $\delta>0$ we can
find $\tilde\epsilon>0$ such that for any ${\bf x}\in\tilde H^{(in)}_l$ and
$|{\bf x}|<\tilde\epsilon$ the trajectories $\Phi_t({\bf x})$ satisfy
$$d(\Phi_t({\bf x}),X)<\delta\mbox{ for any }t\ge0.$$

In a sufficiently small neighbourhood of the origin at the hyperplane
$\tilde H^{(in)}_j$, the map $\psi_j:\tilde H^{(out)}_{j-1}\to\tilde H^{(in)}_j$
is predominantly linear. For any intermediate crossection $\tilde H^{(int)}_j$
between $\tilde H^{(out)}_{j-1}$ and $\tilde H^{(in)}_j$, the induced map
$\tilde H^{(int)}_j\to\tilde H^{(in)}_j$ is also predominantly
linear. Hence we can find a positive $K_j^{(\rm glob)}$ such that
$$d(\Phi_t({\bf x}),X)<K_j^{(\rm glob)}d(\Phi_{j,k}^{(in)}({\bf x}),X)
\hbox{ for any }t,\ t^{(out)}_{j-1,k}<t<t^{(in)}_{j,k}.$$

Now consider the trajectory $\Phi_t({\bf x})$ at the time interval
$t^{(in)}_{j,k}<t<t^{(out)}_{j,k}$. Near $\xi_j$, we project the cycle $X$ and
the trajectory $\Phi_t({\bf x})$ onto the plane $(v,w)$ and onto the orthogonal
hyperplane $({\bf u},{\bf z})$. By $d^{(v,w)}(\cdot,\cdot)$ and
$d^{({\bf u},{\bf z})}(\cdot,\cdot)$ we denote the distances between
the projections onto the $(v,w)$ plane and onto the $({\bf u},{\bf z})$
hyperplane, respectively. Simple algebra (not presented here) attests that
$$d^{(v,w)}(\Phi_t({\bf x}),X)<
(d(\Phi_{j,k}^{(in)}({\bf x}),X))^{\beta_j/(1+\beta_j)},$$
where $\beta_j=c_j/e_j$. The estimate
$$d^{({\bf u},{\bf z})}(\Phi_t({\bf x}),X)<
d(\Phi_{j,k}^{(in)}({\bf x}),X)+d(\Phi_{j,k}^{(out)}({\bf x}),X)$$
follows from (\ref{lmap}). We denote
$$K^{(\rm glob)}=\max_j(K_j^{(\rm glob)}),\ \beta=\min_j(\beta_j).$$

For a given $\delta>0$, let $\delta_1>0$ satisfy
$$K^{(\rm glob)}\delta_1^{\beta/(1+\beta)}<\delta/2,
\ (K^{(\rm glob)}+1)\delta_1<\delta/2.$$
Since $\bf 0$ is asymptotically stable, we can find $\tilde\epsilon>0$ such
that $d(\Phi_{j,k}^{(in)}({\bf x}),X)<\delta_1$ for all $j$ and $k$, provided
${\bf x}\in\tilde H^{(in)}_l$ for some $l$ and $d({\bf x},X)<\tilde\epsilon$.
The estimates
presented above imply that $d(\Phi_t({\bf x}),X)<\delta$ for all $t>0$.

\medskip
Finally, we consider ${\bf x}\notin H^{(in)}_l$ for all $l$. Denote by
$\tilde {\bf x}$ the first intersection of $\Phi_t({\bf x})$ with some
$H^{(in)}_l$. By the arguments
similar to those presented above, at least one of the estimates
$$d(\tilde {\bf x},X)<Kd({\bf x},X),\hbox{ if ${\bf x}$ is near }\kappa_l$$
and
$$d(\tilde {\bf x},X)<\tilde K(d({\bf x},X)+d({\bf x},X)^{\beta/(1+\beta)}),
\hbox{ if ${\bf x}$ is near }\xi_{l-1}$$
holds true. Each of the two inequalities imply that, for $\tilde\epsilon$
defined in the previous paragraph, we can find $\epsilon>0$ such that
$d({\bf x},X)<\epsilon$ implies $d(\tilde {\bf x},X)<\tilde\epsilon$. Hence for
$d({\bf x},X)<\epsilon$ the estimate $d(\Phi_t({\bf x}),X)<\delta$ holds true
for all $t>0$.

\medskip
The proof that $\lim_{t\to\infty}d(\Phi_t({\bf x}),X)=0$ follows from
$\lim_{k\to\infty}\tilde\pi_j^k\tilde g_{(j-1,l)}{\bf x}=0$ is similar, and
we omit it.
\qed

\begin{lemma}\label{le1}
Let $X$ be a type Z heteroclinic cycle. If $X$ is fragmentarily asymptotically
stable, then for any $\delta>0$ and any $j$
\begin{equation}\label{clem1}
\mu^{n-1}(\tilde H^{(in)}_j\cap \cB_{\delta}(X))>0,
\end{equation}
where $\mu^{n-1}$ is the Lebesgue measure in $\R^{n-1}$.
\end{lemma}

\proof
Denote $Q_j=\tilde H^{(in)}_j\cap \cB_{\delta}(X)$. Suppose (\ref{clem1})
is not satisfied for some $j$, i.e.
\begin{equation}\label{assu1}
\mu^{n-1}(Q_j)=0.
\end{equation}
The set $\cB_{\delta}(X)$ can be regarded as the union of segments of
trajectories going from $Q_j$ in the direction of negative $t$ till an
intersection either with $\tilde H^{(in)}_j$ or with the boundary
of $B_{\delta}(X)$ occurs:
$$\cB_{\delta}=\{\Phi_t({\bf x}):\ {\bf x}\in Q_j,
\ t^{\rm final}<t\le 0\hbox{ where }
\Phi_{t^{\rm final}}({\bf x})\in\tilde H^{(in)}_j\hbox{ or }
d(\Phi_{t^{\rm final}}({\bf x}),X)=\delta\}.$$
Denote by $\cB^{(glob)}_{j,\delta}(X)$ the part of $\cB_{\delta}(X)$ bounded by
$\tilde H^{(out)}_j$ and $\tilde H^{(in)}_{j+1}$, and by
$\cB^{(loc)}_{j,\delta}(X)$ the part bounded by $\tilde H^{(in)}_j$ and
$\tilde H^{(out)}_{j+1}$. By linearity of the global map, (\ref{assu1}) implies
$\mu(\cB^{(glob)}_{j-1,\delta}(X))=0$ and
\begin{equation}\label{muB}
\mu^{n-1}(\tilde H^{(out)}_{j-1}\cap \cB_{\delta}(X))=0.
\end{equation}
Since the local map satisfies (\ref{lmap}) and due to (\ref{muB}),
$$\mu(\cB^{(loc)}_{j-1,\delta}(X))=0\hbox{ and }
\mu^{n-1}(\tilde H^{(in)}_{j-1}\cap \cB_{\delta}(X))=0.$$

The same arguments applied $m-1$ times to the sets
$Q_{j-1}=\tilde H^{(in)}_{j-1}\cap\cB_{\delta}(X)$, $Q_{j-2}$, ...,
imply that $\mu(\cB_{\delta}(X))=0$, in contradiction with the statement
of the lemma. Therefore the assumption (\ref{assu1}) is false,
i.e., $\mu^{n-1}(Q_j)>0$ for all $j$.
\qed

\begin{theorem}\label{th2}
Let $\{g_1^m\}$, $g_j:\R^{n_t+1}\to\R^{n_t+1}$, be the collection of maps
associated with a type Z heteroclinic cycle. The cycle is fragmentarily
asymptotically stable, if and only if the fixed point $(w,{\bf z})=\bf 0$
of the collection of maps is fragmentarily asymptotically stable.
\end{theorem}

\proof
By lemma \ref{le1}, for any $j$ (\ref{clem1}) holds true. Therefore
the measure $\mu^{n_t+1}$ (in $\R^{n_t+1}$) of the orthogonal projection
of the set $\tilde H^{(in)}_j\cap\cB_{\delta}(X)$ into the plane $\bf q=0$ is
positive. For a small $\delta$ the collection of maps gives accurate
predictions for trajectories $\Phi_t({\bf x})$,
$\Phi_t({\bf x})\subset B_{\delta}(X)$ for $t>0$, and the coordinates
$w$ and $\bf z$ are independent of $\bf q$. Hence $\bf0$ is a fragmentarily
asymptotically stable fixed point of the collection of maps.

\medskip
The proof, that fragmentary asymptotic stability of $\bf0$ of the collection
of maps $\{g_1^m\}$ implies fragmentary asymptotic stability of $\bf0$
of the collection of maps $\{\tilde g_1^m\}$, is similar to the one
in the proof of theorem \ref{th1} and is omitted.

Let the constants $K^{(\rm glob)}$, $\tilde K$ and $\beta$ be defined as in the
proof of theorem \ref{th1}. Same arguments as employed in this proof imply
that the inequalities
$$d(\Phi_t({\bf x}),X)<K^{(\rm glob)}d(\Phi_{j,k}^{(in)}({\bf x}),X)\hbox{ for any }j,\
k\hbox{ and }t,\ t^{(out)}_{j-1,k}<t<t^{(in)}_{j,k}\hbox{ if }
d(\Phi_{j,k}^{(in)}({\bf x}),X)<\delta;$$
$$d^{(v,w)}(\Phi_t({\bf x}),X)<(d(\Phi_{j,k}^{(in)}({\bf x}),X))^{\beta/(1+\beta)}
\hbox{ for any }j,\ k\hbox{ and }t,\ t^{(in)}_{j,k}<t<t^{(out)}_{j,k}
\hbox{ if }d(\Phi_{j,k}^{(in)}({\bf x}),X)<\delta;$$
$$d^{({\bf u},{\bf z})}(\Phi_t({\bf x}),X)<
d(\Phi_{j,k}^{(in)}({\bf x}),X)+d(\Phi_{j,k}^{(out)}({\bf x}),X)
\hbox{ for any }j,\ k\hbox{ and }t,\ t^{(in)}_{j,k}<t<t^{(out)}_{j,k},$$
$$\hbox{if }d(\Phi_{j,k}^{(in)}({\bf x}),X)<\delta
\hbox{ and }d(\Phi_{j,k}^{(out)}({\bf x}),X)<\delta.$$
hold true for a sufficiently small $\delta>0$.

Defining $\delta_1$ as in the proof of theorem \ref{th1}, we find that
$d(\Phi_t({\bf x}),X)<\delta$ for all $t>0$ provided
${\bf x}\in\tilde H^{(in)}_1$ satisfies
${\bf x}\in\cB_{\delta_1}(\{\tilde g_1^m\},{\bf 0})$.
The proof, that ${\bf x}\in\cB_{\delta_1}(\{\tilde g_1^m\},\bf 0)$
implies $\lim_{t\to\infty}d(\Phi_t({\bf x}),X)=0$, is similar.

Let $\tilde H^{(in)}_{1,s}$ and $\tilde H^{(in)}_{1,-s}$ be two hyperplanes
parallel to $\tilde H^{(in)}_1$ and located at distance $s$ from this
hyperplane. Consider the set $\tilde Q_1$ comprised of pieces of trajectories
contained between the hyperplanes $\tilde H^{(in)}_{1,s}$ and
$\tilde H^{(in)}_{1,-s}$, whose points of intersection with the hyperplane
$\tilde H^{(in)}_1$ constitute the set
$Q_1:=\cB_{\delta_1}(\{\tilde g_1^m\},\bf 0)$. For small $s$,
$\tilde Q_1\subset\cB_{\delta}(X)$ and $\mu(\tilde Q_1)=2s\mu^{n-1}(Q_j)>0$.
Thus the set $X$ is fragmentarily asymptotically stable.
\qed

\section{Stability of fixed points of a collection of maps}\label{smaps}

\subsection{Transition matrix}

Denote by ${\cal M}_j$ the maps $g_j$ in the new coordinates\footnote{Here
we ignore the matrix $A_{\pm}$ which is irrelevant in the study
of stability. It becomes important in the study of bifurcations -- $A_{\pm}$
determines the length of periodic orbit(s) bifurcating from a heteroclinic
cycle (see subsection \ref{bifr}).} \mbel, where
\begin{equation}\label{newc}
\mbe=(\ln|w|,\ln|z_1|,...,\ln|z_{n_t}|).
\end{equation}
As discussed in subsection \ref{cycmap}, the maps are linear, their
structure being
\begin{equation}\label{fmap}
{\cal M}_j\mbe=M_j\mbe+F_j,
\end{equation}
where
\begin{equation}\label{esm}
M_j:=A_jB_j=A_j\left(
\begin{array}{ccccc}
b_{j,1}&0&0&\ldots&0\\
b_{j,2}&1&0&\ldots&0\\
b_{j,3}&0&1&\ldots&0\\
.&.&.&\ldots&.\\
b_{j,N}&0&0&\ldots&0
\end{array}
\right)
\end{equation}
are the {\em basic transition matrices} of the maps. Here $A_j$ and $B_j$ are
$N\times N$ matrices, $N=n_t+1$, $A_j$ is a permutation matrix, and the entries
$b_{j,l}$ of the matrix $B_j$ depend on the eigenvalues of the
linearisation $df(\xi_j)$ of (\ref{eq_ode}) near $\xi_j$ via the relation
\begin{equation}\label{coeB}
b_{j,1}=c_j/e_j\mbox{ and }b_{j,l+1}=-t_{j,l}/e_j,\ 1\le l\le n_t,\ 1\le j\le m.
\end{equation}
We call $\{{\cal M}_1^m\}$, as $\{g_1^m\}$, a collection of
maps associated with the heteroclinic cycle. A fixed point $(w,{\bf z})=\bf 0$
of the collection $\{g_1^m\}$ becomes a fixed point $\mbe=-\mbi$
of the collection $\{{\cal M}_1^m\}$. In the study of stability of the point
$(w,{\bf z})=\bf 0$ we consider asymptotically small $z$ and $\bf w$, i.e.,
asymptotically large negative \mbel, and hence finite $F_j$ can be ignored.

Transition matrices of the superposition of maps $\pi_j$ and $g_{j,l}$
are the products $M^{(j)}=M_{j-1}\ldots M_1M_m\ldots M_{j+1}M_j$ and
$M_{j,l}=M_j\ldots M_1M_m\ldots M_{l+1}M_l$ (or $M_j\ldots M_l$ if $j>l$),
respectively. For a collection of permutation matrices $A_j$ we define
$A^{(j)}$ and $A_{j,l}$ in the same way. Denote by $\lambda_s$ the eigenvalues
of matrices $M^{(j)}$ (they are independent of $j$, since all matrices
$M^{(j)}$ are similar) enumerated in the descending order of their real parts
(generically all the real parts are distinct except for pairs of complex
conjugate eigenvalues). Let also ${\bf w}^{j,s}$ denote the eigenvector
of the matrix $M^{(j)}$ associated with the eigenvalue $\lambda_s$.

\subsection{Two types of eigenvalues of a transition matrix}
\label{twot}

Consider a matrix $M:=M^{(1)}=M_m\ldots M_1:\R^N\to\R^N$; it is
a product of the basic transition matrices of the form (\ref{esm}). We separate
the coordinate vectors ${\bf e}_l$, $1\le l\le N$, into two groups.
The first group is comprised of the vectors ${\bf e}_l$ for which there exist
such $k$ and $j$ that $(A^{(j)})^kA_{j-1,1}{\bf e}_l={\bf e}_1$ (recall that
$A_j$ are permutation matrices), the second one incorporates the remaining
vectors. Denote by $V^{\rm sig}$ and $V^{\rm ins}$ the subspaces spanned
by vectors from the first and second group, respectively (the superscripts
``ins'' and ``sig'' stand for significant and insignificant).

\begin{theorem}\label{th_22}
Let $V^{\rm sig}$ and $V^{\rm ins}$ be the subspaces defined above.

\begin{itemize}
\item[(a)] The subspace $V^{\rm ins}$ is $M$-invariant and the absolute value
of all eigenvalues associated with the eigenvectors from this subspace is one.
\item[(b)] Generically all components of eigenvectors that do not
belong to $V^{\rm ins}$ are non-zero.
\end{itemize}
\end{theorem}

\proof
(a) Denote by $V^{\rm ins}_j$, $2\le j\le m$, the subspaces constructed
for matrices $M^{(j)}$ similarly to $V^{\rm ins}$. Since
${\bf e}_1\notin V^{\rm ins}_j$, the action of $B_j$ on $V^{\rm ins}_j$ is
trivial. Evidently, $A_jV^{\rm ins}_j=V^{\rm ins}_{j+1}$. Hence the actions
of $M$ and $A$ coincide on the subspace $V^{\rm ins}$. The $A$-invariance of
$V^{\rm ins}$ follows from the definition of $V^{\rm ins}$, and hence
$V^{\rm ins}$ is $M$-invariant. The matrix $A$ is a permutation matrix, since
it is a product of permutation matrices, and thus all its eigenvalues have
the unit absolute value. Consequently, the same holds true for the restriction
of $M$ on $V^{\rm ins}$.

\medskip
(b) Let ${\bf w}^{1,s}$ be an eigenvector of $M$ that does not belong to
$V^{\rm ins}$. At least one significant component of ${\bf w}^{1,s}$ does
not vanish. Denote it by $w^{1,s}_l$. By definition of $V^{\rm sig}$, there
exist such $j$ and $k$ that $(A^{(j)})^kA_{j-1,1}{\bf e}_l={\bf e}_1$.
Eigenvectors ${\bf w}^{j,s}$ and ${\bf w}^{1,s}$ are related as follows:
$${\bf w}^{j,s}=(M^{(j)})^kM_{j-1,1}{\bf w}^{1,s}$$
(up to an arbitrary factor). Hence generically the first component
of ${\bf w}^{j,s}$ does not vanish. Since eigenvectors satisfy the relation
$${\bf w}^{1,s}=M_{m,j}{\bf w}^{j,s}$$
and (\ref{esm}) holds true for all matrices $M_{j'}$ in the product $M_{m,j}$,
generically all components of ${\bf w}^{1,s}$ are non-zero.
\qed

We call {\it insignificant} the eigenvalues associated with eigenvectors
from $V^{\rm ins}$, and {\it significant} the rest ones. Generically
the absolute values of all significant eigenvalues differ from one.

\subsection{Two lemmas}\label{twol}

Here we prove two lemmas which will be used in the next subsection to establish
some results on asymptotic stability of a fixed point of a collection
of maps associated with a heteroclinic cycle; as we have proved in section 3,
stability of the fixed point implies asymptotic stability of the cycle.
Arbitrary matrices $M$ are considered in this subsection (except for they are
supposed to have non-negative entries in lemma \ref{lem_a2}); in particular,
they are not assumed to be products of matrices of the form (\ref{esm}).

We consider a criterion for positivity of the measure of the set
$$
U^{-\infty}(M)=\{{\bf y}:\ {\bf y}\in\R^N_-,\ \lim_{k\to\infty}M^k{\bf y}=-\mbi\}
$$
in terms of the dominant eigenvalue and the associated eigenvector
of matrix $M:\R^N\to\R^N$. We have denoted
$$
\R^N_-=\{{\bf y}=(y_1,\ldots,y_N):\ y_j<0\hbox{ for all }j\},\quad
\bar\R^N_-=\{{\bf y}=(y_1,\ldots,y_N):\ y_j\le0\hbox{ for all }j\},
$$
$$
U_S=\{{\bf y}:\ \max_s y_s<S\},\quad\bar U_S=\{{\bf y}:\ \max_s y_s\le S\}.
$$

Upon the change of variables (\ref{newc}), the Lebesgue measure of a set
initially of a finite measure can become infinite. To avoid this, the measure
of a set $V$ in the original variables is regarded as its measure. Note that
$\mu(V)$ is strictly positive, if and only if this is true for the image of $V$
under the mapping ${\bf y}\to{\rm e}^{\bf y}$ inverse to (\ref{newc}).

Denote by $\lambda_{\max}$ and ${\bf w}^{\max}$ the maximum in absolute value
eigenvalue of the matrix $M:\R^N\to\R^N$ and the associated eigenvector,
respectively,
and by $w^{\max}_l$ the components of the eigenvector. If $|\lambda_{\max}|>1$, all
other eigenvalues (except for $\overline{\lambda}_{\max}$ if $\lambda_{\max}$ is
complex) are supposed to be strictly smaller in absolute value than $\lambda_{\max}$.

\begin{lemma}\label{lem_a1}
$\mu(U^{-\infty}(M))$ depends on $\lambda_{\max}$ and ${\bf w}^{\max}$ as follows:
\begin{itemize}
\item[(i)] If $|\lambda_{\max}|\le1$, then $U^{-\infty}(M)=\emptyset$.
\item[(ii)] If $\lambda_{\max}$ is real and
$\lambda_{\max}<-1$, then $\mu(U^{-\infty}(M))=0$.
\item[(iii)] If $\lambda_{\max}$ is complex and $|\lambda_{\max}|>1$, then
$\mu(U^{-\infty}(M))=0$.
\item[(iv)] If $\lambda_{\max}$ is real, $\lambda_{\max}>1$ and
$w_l^{\max}w_q^{\max}\le0$ for some $l$ and $q$, then $\mu(U^{-\infty}(M))=0$.
\item[(v)] If $\lambda_{\max}$ is real, $\lambda_{\max}>1$,
$w_l^{\max}w_q^{\max}>0$ for all $1\le l,q\le N$, then $\mu(U^{-\infty}(M))>0$.
\end{itemize}
\end{lemma}

\proof
Consider the expansion of ${\bf y}\in\R^N_-$
\begin{equation}\label{decy}
{\bf y}=\sum_{i=1}^N a_i{\bf w}_i
\end{equation}
in the basis comprised of eigenvectors of $M$, ${\bf w}_i$, $1\le i\le N$,
and denote by $a_{\max}$ the coefficient in front of ${\bf w}^{\max}$
in the sum (\ref{decy}). Then the $k$-th iterate is
\begin{equation}\label{mn}
M^k{\bf y}=\sum_{i=1}^N\lambda_i^ka_i{\bf w}_i.
\end{equation}
Consequently

\mi
(1) If $|\lambda_{\max}|\le1$, then
$$\lim_{k\to\infty}M^k{\bf y}\ne-\mbi\hbox{ for any }{\bf y}\in\R^N,$$
because if $\lim_{k\to\infty}\lambda_i^ka_i$ exists, it is zero or finite.

\mi
(2)
If $\lambda_{\max}$ is real and $\lambda_{\max}<-1$, then
$$\hbox{ for any }{\bf y}\in\R^N\hbox{ such that }a_{\max}\ne0\hbox{ the limit }
\lim_{k\to\infty}M^k{\bf y}\hbox{ does not exist},$$
because for $k\to\infty$ the iterates $M^k\bf y$ become aligned with
$\lambda_{\max}^k{\bf w}^{\max}$ and the signs of individual components of
$\lambda_{\max}^k{\bf w}^{\max}$ alternate for odd and even $k$.

\mi
(3) If $\lambda_{\max}$ is complex and $|\lambda_{\max}|>1$, then
$$\hbox{ for any }{\bf y}\in\R^N\hbox{ such that }a_{1,\max}\ne0\hbox{ or }
a_{2,\max}\ne0,\hbox{ the limit }\lim_{k\to\infty}M^k{\bf y}
\hbox{ does not exist},$$
because for $k\to\infty$ the iterates $M^k\bf y$ are attracted by the plane
spanned by ${\bf w}^{1,\max}$ and ${\bf w}^{2,\max}$. The map $M$ expressed
in the polar coordinates in the plane amounts to multiplication by
$r{\rm e}^{{\rm i}\psi}$ (i.e. it involves rotation by an angle $\psi\ne2\pi$
and multiplication by $r>1$). However, only one quadrant of the plane belongs
to $\R_-^N$. Hence, the limit does not exist.

\mi
(4) If $\lambda_{\max}$ is real, $\lambda_{\max}>1$ and
$w^{\max}_lw^{\max}_q\le0$ for some $l$ and $q$, then
$$\hbox{ for any }{\bf y}\hbox{ such that }a_{\max}\ne0\
\lim_{k\to\infty}M^k{\bf y}\notin\R^N_-.$$
To show this note that for $k\to\infty$ the iterates $M^k\bf y$ become
asymptotically close to $a_{\max}\lambda_{\max}^k{\bf w}^{\max}$, the signs of
two individual components, $a_{\max}\lambda_{\max}^kw_l^{\max}$ and
$a_{\max}\lambda_{\max}^kw_q^{\max}$, are opposite and hence one of them is
positive (unless one or both of the two components vanish).

\mi
(5) If $\lambda_{\max}$ is real, $\lambda_{\max}>1$ and $w_l^{\max}w_q^{\max}>0$
for all $l$ and $q$ (to be specific, we can assume all $w_l^{\max}>0$), then
$$\lim_{n\to\infty}M^n{\bf y}=-{\mbi}\hbox{ for any }{\bf y}\in\R^N_-
\hbox{ such that }a_{\max}<0,$$
because for $k\to\infty$ the iterates $M^k{\bf y}$ become asymptotically close
to $a_{\max}\lambda_{\max}^k{\bf w}^{\max}$.

\medskip
Clearly, (1)--(4) imply (i)--(iv), respectively.

\medskip
To prove (v), note that the point $-{\bf w}^{\max}$ belongs to $\R^N_-$.
Therefore there exists a neighbourhood $V\subset\R^N_-$ of $-{\bf w}^{\max}$
such that $a_{\max}<0$ for any ${\bf y}\in V$. The measure of this
neighbourhood is positive. Thus (5) implies (v).
\qed

\begin{lemma}\label{lem_a2}
Let $M$ be a matrix with non-negative entries, $|\lambda_{\max}|>1$ and
$w^{\max}_l\ne 0$ for all $1\le l\le N$. Then
\begin{itemize}
\item[(i)] $\lambda_{\max}$ is real and positive.
\item[(ii)] $w_l^{\max}w_q^{\max}>0$ for all $1\le l,q\le N$.
\item[(iii)] $U^{-\infty}(M)=\R^N_-$.
\end{itemize}
\end{lemma}

\proof
Consider ${\bf y}\in\R^N_-$. Since all entries of $M$ are non-negative,
\begin{equation}\label{mkin}
M^k({\bf y})\in\bar\R^N_-\hbox{ for all }k.
\end{equation}
Consider expansion (\ref{decy}) for $\bf y$. If $\lambda_{\max}$ is complex or
real negative, then some iterates $M^k{\bf y}$ are not in $\bar\R^N_-$,
as noted in the proof of lemma \ref{lem_a1}. Similarly, if
$w_l^{\max}w_q^{\max}<0$ for some $l$ and $q$, then (\ref{mkin}) does not
hold true for sufficiently large $k$. No components of the eigenvector
${\bf w}^{\max}$ vanish. Consequently, $w_l^{\max}w_q^{\max}\ne 0$ for all $l$
and $q$, and therefore (ii) holds true.

\medskip
To prove (iii), note that if $a_{\max}>0$ in the expansion (\ref{decy}) for
$\bf y$, then $\lim_{k\to\infty}M^k{\bf y}=\mbi$ which is prohibited by
(\ref{mkin}). Now let us show that $a_{\max}$ in expansion (\ref{decy}) is
non-zero. Suppose $a_{\max}=0$ for some ${\bf y}\in\R^N_-$. There
exists a neighbourhood $U\subset\R^N_-$ of $\bf y$, and hence there exists
$\tilde{\bf y}\in U$ such that in expansion (\ref{decy}) for
$\tilde{\bf y}$ the factor in front of ${\bf w}^{\max}$ is positive,
which contradicts (\ref{mkin}). Hence, $a_{\max}<0$ for all ${\bf y}\in\R^N_-$;
this implies $\lim_{k\to\infty}M^k{\bf y}=-\mbi$.
\qed

\subsection{Properties of maps}

In this subsection we prove two theorems concerning the Poincar\'e maps for
heteroclinic cycles constructed in subsection \ref{cycmap}. These theorems will
be used in investigation of stability of fixed points of a collection of maps.
The Poincar\'e maps ${\cal M}^{(j)}$ are superpositions of maps (\ref{fmap}):
\begin{equation}\label{fmap1}
{\cal M}^{(j)}=
{\cal M}_{j-1}\ldots{\cal M}_1{\cal M}_m\ldots {\cal M}_{j+1}{\cal M}_j.
\end{equation}
In the coordinates $\mbe$ (\ref{newc}), they reduce to
\begin{equation}\label{fmapp}
{\cal M}^{(j)}\mbe=M^{(j)}\mbe+{\bf C}^{(j)}.
\end{equation}
We denote
${\cal M}_{j,l}={\cal M}_j\ldots{\cal M}_1{\cal M}_m\ldots{\cal M}_{l+1}{\cal M}_l$
(or ${\cal M}_j\ldots{\cal M}_l$ if $j>l$).
We will consider matrices $M$ (for instance, $M=M^{(j)}$) that
are products of basic transition matrices of the form (\ref{esm}).

For a linear map $\cal M$, where
\begin{equation}\label{fmap0}
{\cal M}\mbe=M\mbe+{\bf C},
\end{equation}
we define
$$
U^{-\infty}({\cal M})=
\{{\bf y}:\ {\bf y}\in\R^N_-,\ \lim_{k\to\infty}{\cal M}^k{\bf y}=-\mbi\}.
$$

\begin{lemma}\label{lem_2}
Let $\lambda_{\max}$ be the largest in absolute value significant eigenvalue
of the matrix $M$ in (\ref{fmap0}) and ${\bf w}^{\max}$ be the associated
eigenvector. Suppose $\lambda_{\max}\ne 1$ (as noted in subsection \ref{twot},
generically this is true). The measure $\mu(U^{-\infty}({\cal M}))$ is
positive, if and only if the three following conditions are satisfied:
\begin{itemize}
\item[(i)] $\lambda_{\max}$ is real;
\item[(ii)] $\lambda_{\max}>1$;
\item[(iii)] $w_l^{\max}w_q^{\max}>0$ for all $l$ and $q$, $1\le l,q\le N$.
\end{itemize}
\end{lemma}

\proof
For a vector ${\bf y}={y_1,\ldots,y_N}\in\R^N_-$ expressed as a linear
combination (\ref{decy}) of eigenvectors ${\bf w}_i$, we have
$${\cal M}{\bf y}=\sum_{i=1}^N(\lambda_ia_i+d_i){\bf w}_i,$$
where $d_i$ is the component of $\bf C$ in the direction of ${\bf w}_i$.
By simple algebra,
\begin{equation}\label{sumk}
{\cal M}^k{\bf y}=\sum_{i=1;\ \lambda_i\ne1}^N
((a_i-d_i(1-\lambda_i)^{-1})\lambda_i^k+d_i(1-\lambda_i)^{-1}){\bf w}_i+
\sum_{i=1;\ \lambda_i=1}^N(a_i+kd_i){\bf w}_i.
\end{equation}
Suppose that (i)-(iii) hold true and
$a_{\max}-d_{\max}(\lambda_{\max}-1)^{-1}<0$. Thus for $k\to\infty$ the
iterates ${\cal M}^k{\bf y}$ become asymptotically close to
\begin{equation}\label{limcalm}
{\cal M}^k{\bf y}=(a_{\max}-d_{\max}(1-\lambda_{\max})^{-1})
\lambda_{\max}^k{\bf w}^{\max}+{\bf r}_k,
\end{equation}
and for $k\to\infty$ the residual term ${\bf r}_k$ is infinitely small
compared to the first one. Thus $\mu(U^{-\infty}({\cal M}))>0$
by the same arguments as in the proof of part (v) of lemma \ref{lem_a1}.

\medskip
Suppose $|\lambda_{\max}|>1$. For $k\to\infty$, the first term in
(\ref{limcalm}) is again asymptotically the largest one and hence
predominantly determines ${\cal M}^k\bf y$. Therefore, if (i) or (iii)
are not satisfied or $\lambda_{\max}<0$, then $\mu(U^{-\infty}({\cal M}))=0$
by the arguments employed in the proof of lemma \ref{lem_a1}.

Consider now
the case $|\lambda_{\max}|<1$. Suppose that (upon cartesian coordinates are
reordered, if necessary) the insignificant subspace of $M$ consists of vectors\\
${\bf y}={0,\ldots,0,y_{N-n_i+1},\ldots,y_N}\in\R^N$ (the insignificant
subspace is defined in subsection \ref{twot}). We can express
${\cal M}^k{\bf y}={y_{1,k},\ldots,y_{N,k}}$ (\ref{sumk}) as
$${\cal M}^k{\bf y}=
\sum_{i=1;\hbox{ \small significant ${\bf w}_i$}}^N
((a_i-d_i(1-\lambda_i)^{-1})\lambda_i^k+d_i(1-\lambda_i)^{-1}){\bf w}_i$$
$$+\sum_{i=1;\hbox{ \small insignificant ${\bf w}_i$};\ \lambda_i\ne1}^N
((a_i-d_i(1-\lambda_i)^{-1})\lambda_i^k+d_i(1-\lambda_i)^{-1}){\bf w}_i+
\sum_{i=1;\ \lambda_i=1}^N(a_i+kd_i){\bf w}_i.$$
The first sum has a finite limit as $k\to\infty$. In the second and
third sums, $y_{1,k},\ldots,y_{n_s,k}$ vanish as proved in theorem
\ref{th_22}. Therefore $\lim_{k\to\infty}{\cal M}^k{\bf y}\ne-\mbi$.
\qed

Below $\lambda_{\max}\ne 1$ denotes the largest significant eigenvalue of
a transition matrix $M^{(j)}$.

\begin{theorem}\label{th_3}
Let $M_j$ be basic transition matrices of a collection of maps $\{g_1^m\}$
associated with a heteroclinic cycle of type Z. Suppose that for all $j$,
$1\le j\le m$, all transverse eigenvalues of $df(\xi_j)$ are negative. Then
\begin{itemize}
\item[(a)] If the inequality $|\lambda_{\max}|>1$ holds true
for the transition matrix $M:=M^{(1)}=M_m\ldots M_1$, then $\bf 0$ is
an asymptotically stable fixed point of the collection of maps $\{g_1^m\}$.
\item[(b)] If $|\lambda_{\max}|\le1$, then $\bf 0$ is completely unstable.
\end{itemize}
\end{theorem}

\proof
(a) The matrices $M^{(j)}$ are similar, hence if the maximum absolute value
of eigenvalues of matrix $M$ is larger than unity, this is also the case
for $M^{(j)}$ for any $j$. All transverse eigenvalues of $df(\xi_j)$ being
negative, (\ref{esm}) and (\ref{coeB}) imply that all entries of matrices
$M_j$ are non-negative. Since $|\lambda_{\max}|>1$, by theorem \ref{th_22}
$w_q^{\max}\ne 0$ for all $1\le q\le N$. Hence by lemma \ref{lem_a2}
\begin{equation}\label{t1_st1}
\lim_{k\to\infty}(M^{(j)})^kM_{j-1,l}{\bf y}=-\mbi
\end{equation}
for all $1\le j,l\le m$ and any ${\bf y}\in\R^N_-$.

As noted in the proof of lemma \ref{lem_a2}, ${\bf y}\in\R^N_-$ implies
the inequality $a_{\max}<0$ for the coefficient in front of ${\bf w}^{\max}$
in the expansion (\ref{decy}) for $\bf y$. Therefore there exists $\tilde S<0$
such that $a_{\max}-d_{\max}(1-\lambda_{\max})^{-1}<0$ for any
${\bf y}\in U_{\tilde S}$ in (\ref{decy}). Thus, by the same arguments as
employed in the proof of lemma \ref{lem_2},
\begin{equation}\label{t1_st4}
U_{\tilde S}\subset U^{-\infty}({\cal M}).
\end{equation}

Denote by $V^{\perp}(\tilde S)$ the intersection of the set $\bar U_{\tilde S}$
with the $N-1$-dimensional hyperplane orthogonal to the line $(1,\ldots,1)$
and crossing this line at the point $(2\tilde S,\ldots,2\tilde S)$.
The set $V^{\perp}(\tilde S)$ is compact; therefore, by virtue of the inclusion
(\ref{t1_st4}), for any $l$ and $j$ there exists a constant $Q^{l,j}$ such that
\begin{equation}\label{t1_st3}
\max_{1\le s\le N}{\bf e}_s\cdot({\cal M}^{(j)})^k{\cal M}_{j-1,l}{\bf y}<Q^{l,j}
\end{equation}
for all $k>0$ and ${\bf y}\in V^{\perp}(\tilde S)$; here $\{{\bf e}_s\}$
are cartesian coordinate system vectors. Denote
$$Q_{\max}=\max_{1\le l,j\le N}Q^{l,j}.$$

All entries of the matrices $M_j$ are non-negative and the set
$V^{\perp}(\tilde S)$ is compact; hence for any $l$ and $j$ (\ref{t1_st1})
implies existence of a constant $R^{l,j}<0$ such that
\begin{equation}\label{t1_st2}
\max_{1\le s\le N}{\bf e}_s\cdot(M^{(j)})^kM_{j-1,l}{\bf y}<R^{l,j}
\end{equation}
for all $k>0$ and ${\bf y}\in V^{\perp}(\tilde S)$. Denote
$$R_{\max}=\max_{1\le l,j\le N}R^{l,j}.$$

By virtue of (\ref{sumk}),
\begin{equation}\label{sumkn}
{\cal M}^k{\bf y}=M^k({\bf y}-\tilde{\bf y})+\tilde{\bf y}+
\sum_{i=1;\ \lambda_i=1}^Nkd_i{\bf w}_i,
\end{equation}
where
$$\tilde{\bf y}=\sum_{i=1;\ \lambda_i\ne1}^Nd_i(1-\lambda_i)^{-1}{\bf w}_i.$$
For a given $R<0$, set $S=2\tilde S(R+R_{\max}-Q_{\max})/R_{\max}$.
The linearity of maps $M_j$ and (\ref{sumkn}) imply
$$\max_{1\le s\le N}{\bf e}_s\cdot({\cal M}^{(j)})^k{\cal M}_{j-1,l}{\bf y}<R$$
for all $k>0$, $1\le j,l\le m$ and ${\bf y}\in U_S$.

Changing coordinates from {\mbel} back to $(w,\bf z)$, we conclude that by
definition \ref{def8} the point $\bf0$ is an asymptotically stable fixed point
of the collection $\{g_1^m\}$.

\medskip
(b) The statement of the theorem follows from lemma \ref{lem_2}.
\qed

\medskip
The theorem \ref{th_3} is proved in \cite{hs98} for the particular case of a heteroclinic
cycle defined by the replicator equation. This proof relies on the
Perron--Frobenius theorem and can be applied to any heteroclinic cycle
of type Z with an empty insignificant subspace.

\begin{theorem}\label{th_4}
Let $M_j$ be basic transition matrices of a collection of maps $\{g_1^m\}$
associated with a heteroclinic cycle of type Z.
(For type Z heteroclinic cycles the matrices are of the form (\ref{esm}).)
Denote by $j=j_1,\ldots j_L$ the indices, for which $M_j$ involves negative
entries; all entries are non-negative for all remaining $j$. Assume $L>0$
(the case $L=0$ is treated by theorem \ref{th_3}).
\begin{itemize}
\item[(a)] If for at least one $j=j_l+1$
the matrix $M^{(j)}$ does not satisfy conditions (i)-(iii) of lemma \ref{lem_2},
then $\bf 0$ is a completely unstable fixed point of the collection $\{g_1^m\}$.
\item[(b)] If the matrices $M^{(j)}$ satisfy conditions (i)-(iii) of lemma
\ref{lem_2} for all $j$ such that $j=j_l+1$, then $\bf 0$ is a fragmentarily
asymptotically stable fixed point of the collection $\{g_1^m\}$.
\end{itemize}
\end{theorem}

\proof
(a) Matrices $M^{(j)}$ satisfy, or not satisfy, conditions (i)-(ii) of lemma
\ref{lem_2} simultaneously for all $j$, because all $M^{(j)}$ are similar.
Hence, by lemma \ref{lem_2}, $\mu(U^{-\infty}({\cal M}^{(j)}))=0$ for all $j$.
Suppose condition (iii) is not satisfied for some $j=J$. The iterates
$({\cal M}^{(J)})^k{\cal M}_{J-1,1}{\bf y}$ tend to align with ${\bf w}^{\max,J}$
in the limit $k\to\infty$ (see (\ref{sumk})\,). Since
$w^{\max,J}_lw^{\max,J}_q\le0$ for some $l$ and $q$, the iterates escape from
$U_S$ for any $S<0$, when $k$ is sufficiently large. Hence part (a) is proved.

\medskip
(b) Suppose for some $l$ the matrix $M^{(j_l+1)}$ satisfies condition (iii)
of lemma \ref{lem_2}. Matrices $M_j$, $j_l+1\le j\le j_{l+1}-1$, have
non-negative entries, they are of the form (\ref{esm}) and
${\bf w}^{\max,j}=M_{j-1}\ldots M_{j_l+2}M_{j_l+1}{\bf w}^{\max,j_l+1}$.
Thus if $w^{\max,j_l+1}_lw^{\max,j_l+1}_q>0$ for all $l$ and $q$,
$1\le l,q\le m$, then $w^{\max,j}_lw^{\max,j}_q>0$ for all $l,\ q$ and $j$,
$j_l+2\le j\le j_{l+1}$.
Hence, it suffices to check condition (iii) only for $j=j_l+1$, $1\le l\le L$.

Denote ${\bf y}^{\rm stab}=Q{\bf w}^{\max}$. By virtue of (\ref{sumk}),
if $Q$ is negative and sufficiently large in magnitude, then
$$({\cal M}^{(j)})^k{\cal M}_{j-1,1}{\bf y}^{\rm stab}\in\R^N_-
\hbox{ for all $1\le j\le m$ and $k\ge0$}$$
and
$$\lim_{k\to\infty}({\cal M}^{(j)})^k{\cal M}_{j-1,1}{\bf y}^{\rm stab}=-\mbi
\hbox{ for all }1\le j\le m.$$

Let $V^{\rm compl}$ be the $N-1$-dimensional $M$-invariant complement to
${\bf w}^{\max}$ in $\R^N$, and $B_1^{N-1}({\bf 0})$ the unit ball
in $V^{\rm compl}$ centred at $\bf 0$. If ${\bf y}\in B_1^{N-1}({\bf 0})$, then
$$\lim_{k\to\infty}\biggl({1\over\lambda_{\max}}M\biggr)^k{\bf y}=0.$$
Therefore there exists a constant $K_1$ such that
$$|\biggl({1\over\lambda_{\max}}M\biggr)^k{\bf y}|<K_1$$
for all ${\bf y}\in B_1^{N-1}({\bf 0})$ and $k\ge0$.
Similarly there exist constants $K_j$, $j=2,\ldots,N$, such that
$$|\biggl({1\over\lambda_{\max}}M^{(j)}\biggr)^kM_{j-1,1}{\bf y}|<K_j$$
for all ${\bf y}\in B_1^{N-1}({\bf 0})$ and $k>0$. Denote $K_{\max}=\max_jK_j$
and assume that upon normalisation ${\bf w}^{\max,j}$ are unit eigenvectors.
Let $d_{\min}$ be the minimum distance from
$({\cal M}^{(j)})^k{\cal M}_{j-1,1}{\bf y}^{\rm stab}\in\R^N_-$
to the $N-1$-dimensional coordinate hyperplanes $y_p=0$
(the minimum is over all hyperplanes, all $k$ and all $j$).

Denote by $B_{d_{\min}/K_{\max}}^{N-1}({\bf y}^{\rm stab})$ the ball
of radius $d_{\min}/K_{\max}$ centred at ${\bf y}^{\rm stab}$
in the $N-1$-dimensional hyperplane parallel to $V^{\rm compl}$, and by
$C_{d_{\min}/K_{\max}}^{N-1}({\bf y}^{\rm stab})$
the semi-infinite cylinder comprised of rays originating at
$B_{d_{\min}/K_{\max}}^{N-1}({\bf y}^{\rm stab})$ (which is the base of the
cylinder) that are parallel to ${\bf w}^{\max}$ and extend towards $-\mbi$.
Suppose ${\bf y}\in C_{d_{\min}/K_{\max}}^{N-1}({\bf y}^{\rm stab})$; then,
by linearity of maps $M_j$ (and hence of their superpositions) and
by construction of the set $C_{d_{\min}/K_{\max}}^{N-1}({\bf y}^{\rm stab})$,
$$({\cal M}^{(j)})^k{\cal M}_{j-1,1}{\bf y}\in\R^N_-
\hbox{ for all }1\le j\le m\hbox{ and }k>0$$
and
$$\lim_{k\to\infty}({\cal M}^{(j)})^k{\cal M}_{j-1,1}{\bf y}=-\mbi
\hbox{ for all }1\le j\le m.$$

Using the same arguments as employed in the proof of theorem \ref{th_3}, we
can show that for any $R<0$ there exists $S<0$ such that
${\bf y}\in C_{d_{\min}/K_{\max}}^{N-1}({\bf y}^{\rm stab})\cap U_S$ implies
$$({\cal M}^{(j)})^k{\cal M}_{j-1,1}{\bf y}\in U_R\hbox{ for all }1\le j\le m
\hbox{ and }k>0.$$
Since the measure of the set
$C_{d_{\min}/K_{\max}}^{N-1}({\bf y}^{\rm stab})\cap U_S$ is positive,
part (b) is proved.
\qed

\section{Bifurcations of heteroclinic cycles}\label{bifc}

In this section we consider codimension-one bifurcations of type Z
heteroclinic cycles. We assume that the system (\ref{eq_ode}) depends
on a scalar parameter, $\alpha$, i.e. it is
\begin{equation}\label{eq_al}
\dot{\bf x}=f({\bf x},\alpha),\quad f:\R^n\times\R\to\R^n.
\end{equation}
A bifurcation takes place at $\alpha=0$ and the cycle exists at the interval
$[\alpha_-,\alpha_+]$, where $\alpha_-<0$ and $\alpha_+>0$. We do not study
bifurcations occurring when the cycle ceases to exist, which happens, e.g.,
if for some $j$ a contracting, expanding or radial eigenvalue of $df(\xi_j)$
vanishes at $\alpha=0$.

\subsection{Transverse bifurcations}\label{bift}

Suppose a transverse eigenvalue of a steady state $\xi_j$ becomes positive
at $\alpha=0$. We have proved in subsection \ref{cycmap} that in a small
neighbourhood of the cycle the Poincar\'e map can be approximated by
a superposition of local and global maps. Near steady states only the linear
part of $f$ has been taken into account. The approximation is accurate
in a neighbourhood of a steady state provided in this neighbourhood
the linear part is significantly larger than the omitted nonlinear terms
in Taylor's expansion of $f$. As an eigenvalue of $df(\xi_j)$ tends to zero,
the neighbourhood of $\xi_j$, where the approximation is valid, shrinks.
At $\alpha=0$, for which the eigenvalue of $df(\xi_j)$ is zero, the collection
of maps cannot be used to study the behaviour of trajectories in the vicinity
of the cycle, but nevertheless we can still comment on bifurcations
of the cycle. Stability of the cycle can change in a bifurcation of a steady
state $\xi_j$, where a transverse eigenvalue of $df(\xi_j)$ becomes positive
for $\alpha>0$, in the following ways:

\begin{itemize}
\item[(a)] The cycle is asymptotically stable for $\alpha<0$ and fragmentarily
asymptotically stable for $\alpha>0$. This happens, if the conditions
(i)-(iii) of lemma \ref{lem_2} remain satisfied for $\alpha>0$. An example
of such a bifurcation is studied in the paper \cite{dh09}.
\item[(b)] The cycle is asymptotically stable for $\alpha<0$ and completely
unstable for $\alpha>0$. This occurs, if at least one of the conditions
(i)-(iii) is violated for $\alpha>0$. Examples of such bifurcations
of homoclinic cycles are studied in \cite{ckms}.
\item[(c)] The cycle is fragmentarily asymptotically stable for $\alpha<0$ and
remains such for $\alpha>0$ (the conditions (i)-(iii) hold true for $\alpha>0$).
\item[(d)] The cycle is fragmentarily asymptotically stable for $\alpha<0$ and
completely unstable for $\alpha>0$ (at least one of the conditions (i)-(iii)
is violated for $\alpha>0$).
\end{itemize}

\medskip
When a transverse eigenvalue of $df(\xi_j)$ changes its sign, a pair of mutually
symmetric steady states, $\xi_j'$ and $\xi_j''$, emerges in a pitchfork bifurcation.
Denote by $v^{\rm crit}$ eigenvector spanning the eigenspace of $df(\xi_j)$,
where the bifurcation takes place. By lemma \ref{lem0}, the subspace
$L_j\oplus v^{\rm crit}$ is invariant for the dynamical system (\ref{eq_al})
implying that the heteroclinic connection $\xi_j\to\xi_j'$ (and, due to
the symmetry, $\xi_j\to\xi_j''$) exists that is structurally stable.
The type of the global object appearing in the bifurcation depends on whether
there exists another structurally stable heteroclinic connection, namely,
$\xi_j'\to\xi_{j+1}$ (or possibly $\xi_j'\to\xi_{j+l}$ with $l>1$).
If such a connection exists, a new heteroclinic cycle
$(\xi_1,\ldots,\xi_j,\xi_j',\xi_{j+l},\ldots,\xi_m)$ is created in the
bifurcation, as, e.g., in the system considered in \cite{dh09}. If the
connection does not exist, more complex objects can possibly emerge, for
instance, a depth-two heteroclinic cycle involving a connection from
$\xi_j'$ to the original heteroclinic cycle $X$.

Let us mention certain particular examples of bifurcating cycles:

\begin{itemize}
\item If the cycle $X$ is homoclinic (i.e., the equilibria are related
by a symmetry $\gamma$, $\xi_{j+1}=\gamma\xi_j$ and
$v^{\rm crit}=\gamma v^{\rm crit}$), then the cycle
$(\xi_1',\xi_2',\ldots,\xi_m')$ emerges that does not involve any $\xi_j$.

\item Suppose that in the cycle $X$ any $\xi_j$ and $\xi_{j+2}$
are related by a symmetry $\gamma$, i.e., $\xi_{j+2}=\gamma\xi_j$ (this does
happen for homoclinic cycles and can occur in other cases) and
$v^{\rm crit}=\gamma v^{\rm crit}$. Then, assuming that the bifurcation takes
place at an even $j$, the bifurcating cycle is
$(\xi_2',\xi_4',\ldots,\xi_m')$ (note $m$ is even).
\end{itemize}

Such bifurcations were studied in \cite{ckms}.
Note that when the dimension of the transverse eigenspace is higher than one
(this is impossible for cycles of type Z), more complex transverse
bifurcations are possible, for instance, the ones discussed in \cite{ckms99}.

\subsection{Resonance bifurcations}\label{bifr}

In this subsection we assume that for any $j$ none of the eigenvalues of
$df(\xi_j)$ crosses the imaginary axis at the interval $\alpha_-<\alpha<\alpha_+$.
Hence there exists a neighbourhood of the heteroclinic cycle, where
trajectories of the system (\ref{eq_al}) are accurately approximated by the
superposition of maps $g_j$, $1\le j\le m$. Thus bifurcations of the cycles
can be investigated by studying bifurcations of fixed points of the collection
of maps associated with the cycle. Here we explore what happens when
the conditions (i)-(iii) of lemma \ref{lem_2} for fragmentary asymptotic
stability of a fixed point $(w,{\bf z})=0$ of the collection of maps are broken.

In subsection \ref{twot} we have divided eigenvalues and eigenvectors of
a transition matrix $M^{(j)}$ into two groups, which we have called significant
and insignificant. We arrange vectors in the basis in $H^{(in)}_j$ (recall
that the basis consists of eigenvectors of $df(\xi_j)$\,) in such a way
that the first $n_s$ vectors are significant and the last $n_i$
ones are insignificant ($N=n_s+n_i$). The first basis vector is the expanding
eigenvector of $df(\xi_j)$. Significant eigenvalues are the eigenvalues of the
left upper $n_s\times n_s$ submatrix of $M^{(j)}$; their absolute values
generically differ from unity. All components of the associated eigenvectors
generically do not vanish. We order the significant eigenvalues so that
${\rm Re}\lambda_j\ge{\rm Re}\lambda_{j+1}$ for $1\le j<n_s$. Insignificant
eigenvalues are one in absolute value, and the first $n_s$ components
of the associated eigenvectors are zero. The transition matrix $M^{(j)}$
restricted to $V^{\rm ins}$ is a matrix of permutation of the basis vectors.
A permutation is a combination of cyclic permutations. We arrange the
insignificant eigenvectors in such a way that each consequent $n_l$ vectors are
involved in the same cyclic permutation of length $n_l$ and the permutation is
${\bf v}^{N_{l-1}+1}\to{\bf v}^{N_{l-1}+2}\to\ldots\to{\bf v}^{N_l}\to{\bf v}^{N_{l-1}+1}$
where $N_l=n_s+n_1+\ldots+n_l$ and $n_i=n_1+\ldots+n_L$.

Conditions (i) and (ii) of lemma \ref{lem_2} for $\lambda_{\max}$ are now
replaced by the following ones:
\begin{itemize}
\item[(i')] $\lambda_1$ is real and $\lambda_1>1$;
\item[(ii')] $|\lambda_1|>|\lambda_j|$ for any $2\le j\le N$.
\end{itemize}

Denote by {\mbzl} a vector in the basis comprised of eigenvectors of matrix
$M$. As above, we set $M:=M^{(1)}$ (our arguments are applicable for any
$M=M^{(j)}$). The map $\cal M$, which is the superposition (\ref{fmap1})
of maps (\ref{fmap}), defines the iterates
\begin{equation}\label{mp0}
\mbe_{n+1}={\cal M}\mbe_n:=M\mbe_n+{\bf c}.
\end{equation}
In these coordinates the iterates reduce to
\begin{equation}\label{mp1}
\zeta_{j,n+1}=\lambda_j\zeta_{j,n}+d_j
\end{equation}
for a real $\lambda_j$, and
\begin{equation}\label{mp2}
\zeta_{j,n+1}=\alpha_j\zeta_{j,n}+\beta_j\zeta_{j+1,n}+d_j,\
\zeta_{j+1,n+1}=\alpha_j\zeta_{j+1,n}-\beta_j\zeta_{j,n}+d_{j+1}
\end{equation}
for complex $\lambda_j=\alpha_j\pm{\rm i}\beta_j$.

\begin{lemma}\label{lem_21}
Consider the following dynamical systems in $\R$ ((a)-(c)) and $\R^2$ (d):
\begin{itemize}
\item[(a)] $x_{n+1}=\lambda x_n+d$, where $\lambda\ne\pm1$ is real;
\item[(b)] $x_{n+1}=\lambda x_n+d$, where $\lambda=-1$;
\item[(c)] $x_{n+1}=\lambda x_n+d$, where $\lambda=1$;
\item[(d)] $x^1_{n+1}=\alpha x^1_n+\beta x^2_n+d^1$,
$x^2_{n+1}=\alpha x^2_n-\beta x^1_n+d^2$.
\end{itemize}

\noindent
Fixed points and periodic orbits of these systems are:
\begin{itemize}
\item[(a)] $x=d/(1-\lambda)$ and $x=\pm\infty$ (if $\lambda<0$, $x=\pm\infty$
is a period-two orbit). The fixed point $x=d/(1-\lambda)$ is stable for
$|\lambda|<1$, and the fixed points $x=\pm\infty$ are stable for $|\lambda|>1$;
\item[(b)] $x=d/2$ is a fixed point, any real number is a period-two orbit;
\item[(c)] $x=\pm\infty$ are the only fixed points. The fixed point $x=\infty$
is stable for $d>0$ and the fixed point $x=-\infty$ is stable for $d<0$;
\item[(e)] $x^1=(d^1(1-\alpha)+\beta d^2)/((\alpha-1)^2+\beta^2))$,
$x^2=(d^2(1-\alpha)-\beta d^1)/((\alpha-1)^2+\beta^2))$ is a unique fixed point,
stable for $|\lambda|<1$.
\end{itemize}
\end{lemma}

The statements of the lemma are directly verified by a simple algebra.

\begin{theorem}\label{th_6}
Let $M_j$, $1\le j\le N$, be basic transition matrices of the collection
of maps $\{g_1^m\}$ associated with a type Z heteroclinic cycle (this implies
that the transition matrices have the form (\ref{esm})). Suppose
\begin{itemize}
\item[(i)] the entries $b_{j,l}$ of the basic transition matrices depend
continuously on $\alpha$;
\item[(ii)] for $\alpha_-\le\alpha<\alpha_+$ condition (iii)
of lemma \ref{lem_2} is satisfied for all $M^{(j)}$;
\item[(iii)] for $\alpha_-\le\alpha<0$ significant eigenvalues of matrix $M$
satisfy the conditions $\lambda_1>1$ and $|\lambda_j|<1$ for $2\le j\le n_s$;
\item[(iv)] for $0<\alpha\le\alpha_+$ all significant eigenvalues of matrix
$M$ satisfy $|\lambda_j|<1$ for all $1\le j\le n_s$.
\end{itemize}

\noindent
Then
\begin{itemize}
\item[(a)] For $\alpha_-<\alpha<0$ the fixed point $\bf 0$ of the collection of
maps $\{g_1^m\}$ is fragmentarily asymptotically stable, and for $\alpha>0$
it is completely unstable;
\item[(b)] Suppose $d_1<0$ in (\ref{mp1})
%??? and $0<\alpha\le\tilde\alpha$
and the point $\mbe^1$ is defined by the relations
\begin{equation}\label{defeta}
\renewcommand{\arraystretch}{1.6}
\begin{array}{l}
\eta^1_j=\displaystyle{
\sum_{l=1;\lambda_l\hbox{ \small is real}}^{n_s}{d_l\over 1-\lambda_l}w^{1,l}_j
+\sum_{l=1;\lambda_l\hbox{ \small is complex}}^{n_s}
{d_l(1-\rea(\lambda_l))\pm d_{l\pm 1}\ima(\lambda_l)\over |\lambda_l-1|^2}
w^{1,l}_j}\\
\hbox{ for }1\le j\le n_s;\\
\eta^1_j=-\infty\hbox{ for }n_s+1\le j\le N.
\end{array}
\end{equation}
(In the second sum the sign $\pm$ coincides with the sign of the imaginary
part of $\lambda_l$. A component is $-\infty$ whenever in the original
coordinates $(w,{\bf z})$ the respective component vanishes.) Then $\mbe^1$
is an asymptotically stable fixed point of the
collection $\{g_1^m\}$ for $0<\alpha\le\alpha_+$.
Components $\eta^1_j$ (\ref{defeta}) for $1\le j\le n_s$ satisfy
$$\lim_{\alpha\to+0}\eta^1_j=-\infty.$$
\item[(c)] If $d_1>0$, then for $\alpha_-\le\alpha<0$ the point (\ref{defeta})
is an unstable fixed point of the collection $\{g_1^m\}$.
Components $\eta^1_j$ (\ref{defeta}) for $1\le j\le n_s$ satisfy
$$\lim_{\alpha\to-0}\eta^1_j=-\infty.$$
\end{itemize}
\end{theorem}

\proof (a) follows from lemma \ref{lem_2}.

\medskip
(b) Recall that insignificant eigenvectors are arranged in such a way, that
each consequent $n_l$ vectors are involved in the same cyclic permutation.
The subspace spanned by these $n_l$ vectors,
$\{{\bf v}^{1,N_l+1},{\bf v}^{1,N_l+2},\ldots,{\bf v}^{1,N_{l+1}}\}$,
is an invariant subspace of map $M$. The action of the map on this
subspace is determined by the matrix
\begin{equation}\label{matri}
\left(
\begin{array}{ccccc}
0&1&0&\ldots&0\\
0&0&1&\ldots&0\\
.&.&.&\ldots&.\\
0&0&0&\ldots&1\\
1&0&0&\ldots&0
\end{array}
\right).
\end{equation}
The $n_l$ eigenvalues of the restriction of $M$ into this invariant
subspace are roots of unity. One of the eigenvalues is unity, all $n_l$
components of the associated eigenvector are equal.

Consider the map $\cal M$ in the basis comprised of eigenvectors of $M$.
We apply lemma \ref{lem_21} in each eigenspace separately to find
a fixed point of the map. For eigenvalues different from unity, we choose
finite fixed points in the associated eigenspaces. For the eigenvalues 1, we
choose the fixed points $-\infty$ in the associated eigenspaces. The change
of coordinates from {\mbzl} to
{\mbel} yields the desirable expression (\ref{defeta}) for the fixed point.

In the sum (\ref{defeta}), the factor $d_1/(1-\lambda_1)$ is asymptotically
larger than any other factor in all other terms. For $\alpha\to+0$ this factor
tends to infinity, while others have finite limits. Hence the components
$\eta^1_j$ for $1\le j\le n_s$ tend to $-\infty$ for $\alpha\to+0$, as required.

Next we prove that the fixed point (\ref{defeta}) is asymptotically stable.
Suppose we know the first $n_s$ components, $(w^{1,1}_1,\ldots,w^{1,1}_{n_s})$,
of the eigenvector ${\bf w}^{1,1}$ (they can be found by calculating the
eigenvectors of the $n_s\times n_s$ upper left submatrix of $M$). The last $n_i$
components of ${\bf w}^{1,1}$ can be obtained from the equations
\begin{equation}\label{syseq}
\renewcommand{\arraystretch}{1.6}
\begin{array}{l}
\sum_{q=1}^{n_s}a_{N_{l-1}+1,q}w^{1,1}_q+w^{1,1}_{N_{l-1}+2}=
\lambda_1w^{1,1}_{N_{l-1}+1}\\
\sum_{q=1}^{n_s}a_{N_{l-1}+2,q}w^{1,1}_q+w^{1,1}_{N_{l-1}+3}=
\lambda_1w^{1,1}_{N_{l-1}+2}\\
\ldots\\
\sum_{q=1}^{n_s}a_{N_{l-1}+n_l,q}w^1_q+w^{1,1}_{N_{l-1}+1}=
\lambda_1w^{1,1}_{N_l},
\end{array}
\end{equation}
where $a_{s,q}$ are the entries of the matrix $M$ (a separate system of
dimension $n_l$ is obtained for each invariant insignificant subspace of $M$).

We solve the system (\ref{syseq}) as follows. Multiplying the first equation by
$\lambda_1^{n_l-1}$, the second one by $\lambda_1^{n_l-2}$, etc. ,
and adding up the resultant equations we obtain
\begin{equation}\label{firw}
w^{1,1}_{N_{l-1}+1}={1\over \lambda_1^{n_l}-1}
\sum_{j=1}^{n_l}\lambda_1^{n_l-j} \sum_{q=1}^{n_s}a_{N_{l-1}+j,q}w^{1,1}_q.
\end{equation}
We proceed similarly to find $w^{1,1}_j$, $N_{l-1}+1<j\le N_l$.

Since $\lambda_1$ tends to unity for $\alpha\to-0$, the condition
$w^{1,1}_j>0$ for $N_{l-1}+1\le j\le N_l$ in this limit is equivalent to
\begin{equation}\label{conb}
\sum_{j=1}^{n_l}\sum_{q=1}^{n_s}a_{N_{l-1}+j,q}w^{1,1}_q>0.
\end{equation}

Stability of the point (\ref{defeta}) to a perturbation from the significant
subspace of $M$ follows from lemma \ref{lem_21} and the fact that all eigenvalues
in the significant subspace are less than one in absolute value. For $\alpha$
close to $+0$, the fixed point $\mbe^1$ has the asymptotics
$C(\alpha){\bf w}^{1,1}$, where $C(\alpha)<0$ is a large constant.

Consider now a perturbation from an invariant subspace of dimension $n_l$
of the insignificant subspace. Choose a positive $D$ such that
$$D>-\min_j\sum_{q=1}^{n_s}a_{N_{l-1}+j,q}w^{1,1}_q.$$
For a given $S<0$, set $R=S-n_lC(\alpha)D$. An initial perturbation ${\bf v}_1$
from this subspace evolves according to the following equations:
\begin{equation}
%\label{syseq}
\renewcommand{\arraystretch}{1.6}
\begin{array}{l}
v^{N_{l-1}+1}_{n+1}=
C(\alpha)\sum_{q=1}^{n_s}a_{N_{l-1}+1,q}w^{1,1}_q+v^{N_{l-1}+2}_n\\
v^{N_{l-1}+2}_{n+1}=
C(\alpha)\sum_{q=1}^{n_s}a_{N_{l-1}+2,q}w^{1,1}_q+v^{N_{l-1}+3}_n\\
\ldots\\
v^{N_l}_{n+1}=C(\alpha)\sum_{q=1}^{n_s}a_{N_l,q}w^{1,1}_q+v^{N_{l-1}+1}_n.
\end{array}
\end{equation}
Due to our choice of $R$ and (\ref{conb}), if initially all components of
${\bf v}_1$ satisfy $v^j_1<R$ for $N_{l-1}+1\le j\le N_l$, then for any $k\ge 1$
the estimate $v^j_k<S$ for $N_{l-1}+1\le j\le N_l$ holds true. Hence, the point
(\ref{defeta}) is stable in the insignificant subspace.

To prove (c), note that expression (\ref{defeta}) still
defines a fixed point of the collection of maps. It bifurcates from $-\mbi$
at $\alpha=0$ and remains close to $-\mbi$ for negative $\alpha$ (recall that
$d_1>0$). The point is unstable in the direction of ${\bf w}^{1,1}$, because
the associated eigenvalue is larger than one.
\qed

Note that if a fixed point of a collection of maps associated with
a heteroclinic cycle exists near $\bf 0$, then there exists a periodic orbit
close to the cycle. Stability (or instability) of the periodic orbit follows
from stability (or instability, respectively) of the fixed point of the
collection of maps. The distance of the bifurcating fixed point from $\bf 0$
is $O(\exp(D/(\lambda_1-1)))$. If the eigenvalue depends linearly on $\alpha$
(the generic case), the distance from the periodic orbit to the heteroclinic
cycle in $H_j^{(in)}$ is proportional to ${\rm e}^{-d/|\alpha|}$, and hence for
any $j$ the time required for the orbit to get from $H_j^{(in)}$ to $H_j^{(out)}$
is proportional to $1/|\alpha|$. Therefore near the point of bifurcation
the temporal period of the bifurcating periodic orbit behaves as $1/|\alpha|$.

In the study of stability we have ignored matrices $A^{\pm}_j$, because only
the distance to $\bf 0$ matters for stability. Denote
$A^{\pm}=A^{\pm}_m\ldots A^{\pm}_1$. Since the matrices $A^{\pm}_j$ are
diagonal, the product is independent of the order of multiplication of the
matrices. Suppose a bifurcating periodic orbit intersects $H_j^{(in)}$ at
a point $x$. The next intersection is at the point $A^{\pm}x$ coinciding with
$x$, if the matrix $A^{\pm}$ is the identity, or is a different point otherwise.
Therefore the bifurcating orbit can have the same length\footnote{We use
the term length in a loose sense here. The length of a heteroclinic
cycle and of the bifurcating periodic orbit can be measured as a number of
connecting trajectories comprising the cycle. Alternatively, it can be
just the Lebesgue measure $\mu^1$ of a one-dimensional set.} as the heteroclinic
cycle under consideration or a twice larger length, depending on the signs
of the diagonal entries of matrix $A^{\pm}$.

In the next theorem we consider bifurcations occurring when $|\lambda_1|>1$,
but one or more conditions of lemma \ref{lem_2} are broken at the point
of bifurcation. In this case the fixed point $\bf 0$ of the collection of maps
$\{g_1^m\}$ ceases to be fragmentarily asymptotically stable at the point
of bifurcation and no new objects (fixed points or invariant sets of other
types) bifurcate from it. Thus, the respective heteroclinic cycle looses
fragmentary asymptotic stability without emergence of bifurcating objects.

If the entries of a matrix depend on a single parameter, generically only two real
eigenvalues of the matrix can become equal as the parameter is varied, and at
the critical parameter value the restriction of the matrix onto the eigenspace
associated with this eigenvalue is a Jordan cell. On further variation
of the parameter the two eigenvalues in this two-dimensional subspace change
into a pair of complex conjugate ones.

\begin{theorem}\label{th_7}
Let $M_j$, $1\le j\le m$, be basic transition matrices of the collection of
maps $\{g_1^m\}$ associated with a type Z heteroclinic cycle. (In particular,
$M_j$ have the form (\ref{esm}).) Suppose
\begin{itemize}
\item[(i)] the entries $b_{j,l}$ of the transition matrices depend
continuously on $\alpha$;
\item[(ii)] for $\alpha_-\le\alpha\le\alpha_+$ the inequalities
$|\lambda_1|>1$ and $\lambda_j\ne 1$ for $1<j\le n_s$ are satisfied;
\item[(iii)] for $\alpha_-\le\alpha<0$ all conditions of lemma \ref{lem_2}
on eigenvectors and eigenvalues of matrices $M^{(j)}$ are satisfied;
\item[(iv)] for $0<\alpha\le\alpha_+$ some conditions of lemma \ref{lem_2}
are not satisfied, i.e. at least one of the following possibilities is realised:
\begin{itemize}
\item[(iv.1)] $\lambda_1$ is complex, $\lambda_1=\bar\lambda_2$ and
$\lim_{\alpha\to+0}\ima(\lambda_1)=\lim_{\alpha\to+0}\ima(\lambda_2)=0$;
\item[(iv.2)] there exists $\lambda_j<0$ such that $|\lambda_j|>\lambda_1$ and
$\lim_{\alpha\to+0}|\lambda_j|=\lambda_1$;
\item[(iv.3)] two eigenvalues $\lambda_j$ and $\lambda_{j+1}=\bar\lambda_j$
are complex, $|\lambda_j|>\lambda_1$ and $\lim_{\alpha\to+0}|\lambda_j|=\lambda_1$;
\item[(iv.4)] there exists $q$ such that $w^{1,1}_q<0$, $w^{1,1}_l>0$ for
$l\ne q$ and $\lim_{\alpha\to+0}w^{1,1}_q=0$.
\end{itemize}
\end{itemize}

\noindent
Then

\begin{itemize}
\item[(a)] For $\alpha_-<\alpha<0$ the fixed point $\bf 0$ of the collection
of maps $\{g_1^m\}$ is fragmentarily asymptotically stable, and for $0<\alpha<\alpha_+$
it is completely unstable.
\item[(b)] There exists a negative $S$ such that the map ${\cal M}:={\cal M}^{(1)}$
does not have fixed points in $U_S$ (other than -\mbil)
for any $\alpha_-<\alpha<\alpha_+$.
\item[(c)] There exists a negative $S$ such that for any ${\bf y}\in U_S$
one of the following statements is true:\\
-- there exist $K>0$ and $h>1$ such that
$|{\cal M}^{k+1}{\bf y}|>h|{\cal M}^k{\bf y}|$ for
any $k>K$\\
or\\
-- for any $K>0$ there exists $k>K$ such that ${\cal M}^k{\bf y}\notin U_S$.\\
(In case (c) the map $\cal M$ has no periodic orbits or other invariant sets
in $U_S$ other than -\mbil.)
\end{itemize}
\end{theorem}

\proof (a) follows from lemma \ref{lem_2}.

\medskip
To prove (b), note that by lemma \ref{lem_21} the first $n_s$ coordinates of
any fixed point $\mbe^1$ of the map $\cal M$ are -\mbil, or otherwise they are
given by (\ref{defeta}). For $S<0$ satisfying
\begin{equation}\label{chS}
S<\min_{1\le j\le n_s,\ \alpha_-<\alpha<\alpha_+}\eta^1_j(\alpha),
\end{equation}
the map $M$ has no fixed points in $U_S$ other than -\mbil. (The values of
$\eta^1_j$, $1\le j\le n_s$, are bounded for $\alpha_-<\alpha<\alpha_+$,
because no significant $\lambda_j$ is equal to 1 for such $\alpha$.)

\medskip
To prove (c), set $S$ as above. Define $\tilde\mbe^1$ by relations
\begin{equation}\label{defetat}
\renewcommand{\arraystretch}{1.6}
\begin{array}{l}
\tilde\eta^1_j=\displaystyle{
\sum_{l=1;\lambda_l\hbox{ \small is real}}^N{d_l\over 1-\lambda_l}w^{1,l}_j+
\sum_{l=1;\lambda_l\hbox{ \small is complex}}^N
{d_l(1-\rea(\lambda_l))\pm d_{l\pm 1}\ima(\lambda_l)\over |\lambda_l-1|^2}
w^{1,l}_j}\\
\hbox{ for }1\le j\le n_s,\\
\tilde\eta^1_j=0
\hbox{ for }n_s+1\le j\le N.\\
\end{array}
\end{equation}
Let ${\bf y}\in U_S$ be decomposed in the basis comprised of eigenvectors of
$M$, with the origin shifted to $\tilde\mbe^1$. First, we note that at least
one of the first $n_s$ coefficients $\tilde\zeta_j$ in the sum
$${\bf y}=\sum_{q=1}^N\tilde\zeta_j{\bf w}^{1,j}$$
is non-zero (otherwise by (\ref{chS}) the point $\bf y$ would be outside $U_S$).
Let $\tilde\lambda_{\max}$ be the largest in absolute value significant
eigenvalue associated with a non-vanishing coefficient. For $k\to\infty$,
the iterates $M^k{\bf y}$ tend to be aligned with the associated eigenvector
$\tilde{\bf w}_{\max}$ (or with the respective two-dimensional eigenspace,
if $\tilde\lambda_{\max}$ is complex). If $\tilde\lambda_{\max}$ is real and
$\tilde\lambda_{\max}>1$, then for large $k$ the iterates $M^k{\bf y}$ behave as
$(\tilde\lambda_{\max})^k\tilde{\bf w}_{\max}$; hence depending on the signs
of components of $\tilde{\bf w}_{\max}$ they either tend to {-\mbil} or escape
from $U_S$. If $|\tilde\lambda_{\max}|<1$, then the iterates $M^k{\bf y}$
are attracted by (\ref{defeta}), which is outside $U_S$ by our choice of $S$.
If $\tilde\lambda_{\max}$ is complex, or if it is real and $\tilde\lambda_{\max}<-1$,
then for any $K>0$ there exists $M^k\bf y$ outside $\R^N_-$ for some $k>K$.
\qed

The bifurcation considered in theorem \ref{th_7} was studied by Postlethwaite
\cite{pos10} for a particular dynamical system in $\R^4$. She proved
that ``there are no dynamical structures which merge with the cycle at the
point of stability loss'', in agreement with our theorem.

\section{Homoclinic cycles}\label{homoc}

\subsection{Transition matrices}

A homoclinic cycle is a heteroclinic cycle, where all equilibria are
related by a symmetry $\gamma$, $\gamma\xi_j=\xi_{j+1}$.
The transition matrix of a homoclinic cycle is a
basic transition matrix (\ref{esm}) which is a product of a permutation
matrix $A$ and a local matrix $B$. Any permutation is a
composition of cyclic permutations. Suppose the permutation defined by matrix
$A$ is a combination of $L+1$ cyclic permutations. The first permutation
involves $n_s$ significant basis vectors, and the last $n_i$ basis vectors are
insignificant. We order basis vectors in agreement with the relations
\begin{equation}\label{perm}
A{\bf e}_j=
\left\{
\renewcommand{\arraystretch}{1.2}
\begin{array}{ll}
{\bf e}_{j+1}, & j\ne N_l\mbox{ for any } 1\le l\le L\\
{\bf e}_{j-n_l+1}, & j=N_l
\end{array}\right.,
\end{equation}
where, as above, we have denoted $N_l=n_s+n_1+n_2+\ldots+n_l$. Without any loss
of genericity we assume that ${\bf e}_1$ is the expanding eigenvector of $df(\xi)$.

For the basis vectors ordered according to (\ref{perm}), the permutation
matrix $A$ is comprised of $L$ non-vanishing diagonal blocks $n_l\times n_l$,
each being of the form (\ref{matri}). Thus the transition matrix is
\begin{equation}\label{tra1}
M=\left(
\begin{array}{cccc|cccc|c}
b_2&1&0&\ldots&0&0&0&\ldots&\ldots\\
b_3&0&1&\ldots&0&0&0&\ldots&\ldots\\
  .&.&.&\ldots&0&0&0&\ldots&\ldots\\
b_1&0&0&\ldots&0&0&0&\ldots&\ldots\\
\hline
b_{n_1+2}&0&0&\ldots& 0&1&0&\ldots&\ldots\\
b_{n_1+3}&0&0&\ldots& 0&0&1&\ldots&\ldots\\
        .&.&.&\ldots& .&.&.&\ldots&\ldots\\
b_{n_1+1}&0&0&\ldots& 1&0&0&\ldots&\ldots\\
\hline
       .&.&.&\ldots& .&.&.&\ldots&\ldots\\
       .&.&.&\ldots& .&.&.&\ldots&\ldots
\end{array}
\right).
\end{equation}

\subsection{Two examples of type Z homoclinic cycles}\label{exa}

In this subsection we present two general examples of type Z homoclinic cycles.
In both cases, our constructions employ a known homoclinic
cycle in $\R^n$ with an empty insignificant subspace (i.e., all eigenvalues
are significant and the permutation matrix $A$ is just a single cyclic
permutation involving $n_s$ vectors). We add $n_1$ insignificant transverse
directions, where the acceptable $n_1$ depends on $n$. This step
(enlargement of the insignificant subspace by adding new dimensions) can be
repeated any number of times.

\medskip
\noindent
{\bf The first example.} Postlethwaite and Dawes \cite{pd10} presented
an example of the system (\ref{eq_ode}) with a $\Z_n\ltimes\Z^n_2$ symmetry
group possessing a type Z homoclinic cycle. The subgroup $\Z^n_2$ acts
on $\R^n$ as $n$ reflections
$$s_j(x_1,\ldots,x_j,\ldots,x_n)=(x_1,\ldots,-x_j,\ldots,x_n);$$
the subgroup $\Z_n$ is generated by the cyclic permutation
$$\rho(x_1,x_2,\ldots,x_n)=(x_n,x_1,\ldots,x_{n-1}).$$
Suppose\\
-- the system (\ref{eq_ode}) has $n$ equilibria $\xi_j$ related by
the symmetry $\rho$, $\rho\xi_j=\xi_{j+1}$; each equilibrium has an isotropy
subgroup $\Z_2^{n_r+2}$;\\
-- the unstable manifold of $\xi_j$ is one-dimensional, has the isotropy
subgroup $\Z_2^{n_r+1}$ and is attracted by $\xi_{j+1}$.\\
Under these assumptions the system has a heteroclinic cycle. The transition
matrix of the cycle is a block of size $(n_t+1)\times(n_t+1)$
(recall that $n=n_t+n_r+2$). Note that instead of the condition that
the unstable manifold of $\xi_j$ is one-dimensional we can impose the weaker
condition that the connection $\kappa_j=W^u(\xi_j)\cap W^s(\xi_{j+1})$ is
one-dimensional and it belongs to $\Fix(\Z_2^{n_r+1})$.

We can enlarge the dimension of (\ref{eq_ode}) by any $K$ dividing $n$
an arbitrary number of times as follows. Consider the action of the
group $\Z_n\ltimes\Z^{n+K}_2$, where the subgroup $\Z^{n+K}_2$ acts on
$\R^{n+K}$ as $n+K$ reflections
$$s_j(x_1,\ldots,x_j,\ldots,x_{n+K})=(x_1,\ldots,-x_j,\ldots,x_{n+K})$$
and the subgroup $\Z_n$ is generated by a combination of cyclic permutations
\begin{equation}\label{perm2}
\rho(x_1,x_2,\ldots,x_n,x_{n+1},x_{n+2},\ldots,x_{n+K})=
(x_n,x_1,\ldots,x_{n-1},x_{n+K},x_{n+1},\ldots,x_{n+K-1}).
\end{equation}
As a result of this modification the size of the transition matrix increases
to $(n_t+1+K)\times(n_t+1+K)$ and a diagonal block (\ref{matri}) of
size $K\times K$ is added to the matrix $A$.

A general third-order system in $\R^{n+K}$ with the above symmetry group is as
follows:
$$\dot x_1=x_1(\nu_1+\sum_{l=1}^{n+K}c_lx_l^2),$$
$$\dot x_{n+1}=x_{n+1}(\nu_2+\sum_{l=1}^{K}d_l\sum_{s=0}^{n/K-1}x_{l+s}^2+
\sum_{l=1}^{K}d_{n+l}x_{n+l}^2).$$
Equations for other $\dot x_j$ are obtained by application of the symmetry
$\rho$. If $\nu_1c_1<0$, the system has $n$ equilibria on coordinate axes,
related by the symmetry $\rho$. Sufficient conditions on $\nu_1$ and $c_l$,
$1\le l\le n$, for existence in $\R^n$ of a homoclinic cycle connecting
the equilibria are presented in \cite{pd10}. Our construction thus implies
existence of a homoclinic cycle in the extended $(n+K)$-dimensional space
for all values of $\nu_2$, $c_l$, $n+1\le l\le n+K$ and $d_l$.

{\bf The second example} follows \cite{dh09}. The authors consider a
heteroclinic cycle in $\R^3$ \cite{agh88,pj98} in a system (\ref{eq_ode})
with the symmetry group $\Z^2\ltimes\Z_2^2$ generated by the symmetries
$$s_1(x_1,x_2,x_3)=(x_1,x_2,-x_3),$$
$$s_2(x_1,x_2,x_3)=(x_1,-x_2,x_3),$$
$$\rho(x_1,x_2,x_3)=(-x_1,x_3,x_2).$$

In \cite{dh09}, the dimension of the system is increased to 5 and
the symmetry group is enlarged to $\Z^4\ltimes\Z_2^3$ by adding the symmetry
\begin{equation}\label{minu}
s_3(x_1,x_2,x_3,x_4,x_5)=(x_1,x_2,x_3,x_4,-x_5).
\end{equation}
The symmetries $s_1$ and $s_2$ act on the added dimensions $(x_4,x_5)$
trivially, and $\rho$ is modified to become
$$\rho(x_1,x_2,x_3,x_4,x_5)=(-x_1,x_3,x_2,-x_5,x_4).$$
The system in $\R^5$ with the above symmetry group truncated at the
third order is as follows:
\begin{equation}\label{sus5}
\renewcommand{\arraystretch}{2.}
\begin{array}{lll}
\dot x_1&=&\nu_1 x_1+c_1(x_2^2-x_3^2)+c_2(x_4^2-x_5^2)+
x_1(c_3x_1^2+c_4(x_2^2+x_3^2)+c_5(x_4^2+x_5^2))\\
\dot x_2&=&x_2(\nu_2+c_6x_1+c_7x_1^2+c_8x_2^2+c_9x_3^2
+c_{10}x_4^2+c_{11}x_5^2)\\
\dot x_3&=&x_3(\nu_2-c_6x_1+c_7x_1^2+c_8x_3^2+c_9x_2^2
+c_{10}x_5^2+c_{11}x_4^2)\\
\dot x_4&=&x_4(\nu_3+d_1x_1+d_2x_1^2+d_3x_2^2+d_4x_3^2+d_5x_4^2+d_6x_5^2)\\
\dot x_5&=&x_5(\nu_3-d_1x_1+d_2x_1^2+d_3x_3^2+d_4x_2^2+d_5x_5^2+d_6x_4^2).
\end{array}
\end{equation}
If $\nu_1c_3<0$, the system possesses two equilibria on the $x$ axis related by
the symmetry $\rho$. Sufficient conditions for existence of a homoclinic cycle
in $\R^3$ involving the equilibria are presented in \cite{agh88,pj98}.
In \cite{dh09} this homoclinic cycle in $\R^5$ is studied for particular values
of coefficients $\nu$, $c$ and $d$ in (\ref{sus5}).

By the same procedure dimension of (\ref{eq_ode}) can be incremented by 1 or 2
any number of times. The dimension of the system can be also expanded by choosing
the dimension of $L_j$ larger than unity, as in the first example considered
in this subsection (the resultant transition matrix is not modified).

\subsection{Stability of a cycle when all transverse eigenvalues are negative}
\label{stabneg}

If all transverse eigenvalues of $df(\xi)$ are negative, the cycle can be
completely unstable or asymptotically stable (see theorem \ref{th_3}). Suppose
vectors comprising the basis are ordered in accordance with (\ref{perm}).
The eigenvalues associated with eigenvectors from the insignificant subspace
are 1 in absolute value (see subsection \ref{twot}). As proved in \cite{pd10},
the dominant eigenvalue of the left upper $n_s\times n_s$ submatrix is real and
larger than one if and only if
\begin{equation}\label{cond1}
b_1+b_2+\ldots+b_{n_s}>1.
\end{equation}
By theorem \ref{th1}, condition (\ref{cond1}) is necessary and sufficient
for a homoclinic cycle of type Z, whose all transverse eigenvalues are
negative, to be asymptotically stable.

\subsection{Two simple cases}
\label{simcas}

If the dimension of the significant subspace is one or two, then the dominant
eigenvalue and the associated eigenvector can be explicitly calculated in
terms of eigenvalues of $df(\xi)$. The left upper submatrix is then
$$
(b_1)\hbox{ \ or \ }
\left(
\begin{array}{cc}
b_2&1\\
b_1&0
\end{array}
\right)
$$
for the dimension one or two, respectively.

If the dimension is one, the eigenvalue is
$$\lambda=b_1$$
and the necessary condition for stability is $b_1>1$.
If the dimension is two, the dominant eigenvalue is
$$\lambda={b_2+\sqrt{b_2^2-4b_1}\over 2},$$
the necessary conditions for stability are $b_2>0$ and $b_1+b_2>1$ (see \cite{pa11}).
These conditions imply that the first two components of the associated
eigenvector have the same sign. We assume that they are positive.

Let us calculate other components of the eigenvector $\bf w$ of the
transition matrix, associated with the dominant eigenvalue $\lambda$. (For
simplicity, the upper indices are omitted.) Consider the eigenvector components
$(w_{N_{l-1}+1},w_{N_{l-1}+2},\ldots,w_{N_l})$
related to the same permutation cycle of length $n_l$. The components
satisfy the equations
$$b_{N_{l-1}+1}w_1+w_{N_{l-1}+2}=\lambda w_{N_{l-1}+1},$$
$$b_{N_{l-1}+2}w_1+w_{N_{l-1}+3}=\lambda w_{N_{l-1}+2},$$
$$\ldots$$
$$b_{N_l}w_1+w_{N_{l-1}+1}=\lambda w_{N_l}.$$
Denote $h(k,n_l)=\mod_{n_l}(k)$ and suppose $0\le k\le n_l-1$. Multiplying
the equations by $\lambda^{h(k+1,n_l)},\ldots,\lambda^{h(k+n_l,n_l)}$,
respectively, and adding them up we obtain
$$(b_{N_{l-1}+1}\lambda^{h(k+1,n_l)}+\ldots+
b_{N_l}\lambda^{h(k+n_l,n_l)})w_1=(\lambda^{n_l}-1)w_{N_{l-1}+k}.$$
Thus the components $w_{N_{l-1}+1},\ldots,w_{N_l}$ all have the same sign
as $w_1$ as long as
\begin{equation}\label{cond4}
b_{N_{l-1}+1}\lambda^{h(k+1,n_l)}+\ldots+
b_{N_l}\lambda^{h(k+n_l,n_l)}>0\hbox{ for any }0\le k\le n_l-1.
\end{equation}

If the length of the first permutation cycle is larger than two, the same
conditions guarantee stability of the cycle, namely, conditions (i)-(iii) of
lemma \ref{lem_2} for eigenvalues and eigenvectors of the significant
$n_s\times n_s$ submatrix, and condition (\ref{cond4}) for each insignificant
permutation cycle. However we cannot express the eigenvalue $\lambda$ in terms
of eigenvalues of $df(\xi)$ except for $n_s=3$ and 4, but in these cases
the expressions are too complex to be of any practical use.

\section{Conclusion}\label{sec_conc}

We have defined type Z heteroclinic cycles as simple cycles with a certain
action of subgroups of the symmetry group of the dynamical system. We have
introduced the notion of fragmentary asymptotic stability of an invariant set
and have derived necessary and sufficient conditions for asymptotic stability
or fragmentary asymptotic stability of type Z heteroclinic cycles. For a type Z
cycle we have calculated the basic transition matrices; the matrices depend
on eigenvalues of the linearisations near steady states comprising the cycle
and the action of the system symmetry group.

If all transverse eigenvalues of the linearisations near steady states are
negative, such a heteroclinic cycle can be either asymptotically stable
or completely unstable. The stability depends on whether the
largest in absolute value eigenvalue of transition matrices (which are products
of the basic transition matrices; all transition matrices are similar) is
larger than one in absolute value. For such type Z cycles, eigenvalues of
transition matrices play the same role, as eigenvalues of the linearisations
in the study of stability of steady states and eigenvalues of the linearisations
of the Poincar\'e maps in the study of stability of periodic orbits.

A type Z cycle, some of whose transverse eigenvalues are positive, can be
fragmentarily asymptotically stable. We have derived a criterion for fragmentary
asymptotic stability in terms of eigenvalues and eigenvectors of its transition
matrices.

We have studied bifurcations occurring when conditions for fragmentary
asymptotic stability are broken. Two types of bifurcations have been identified,
transverse and resonance ones. A detailed study of transverse bifurcations is
problematic, because the collection of maps that we use to study stability
does not approximate correctly trajectories of the system for the critical
parameter value. Nevertheless we have commented on how the stability of a cycle
can change in a bifurcation of a steady state, where a transverse eigenvalue
of the linearisation near a steady state vanishes. For a resonance bifurcation,
we prove that either a periodic orbit is born in it (if the dominant eigenvalue
of a transition matrix becomes smaller than one), or no new invariant objects
emerge in it (if other conditions for fragmentary asymptotic stability are
broken).

We anticipate the following continuations of our study.

Apparently, transition matrices can be used to study stability of simple cycles
of other kinds, although it is not evident how to define a transition matrix
for more complex cycles.

We have not considered all issues related to asymptotic stability and
bifurcations of type Z heteroclinic cycles. As we have noted in section
\ref{bifc}, bifurcations resulting in destruction of the cycle are yet to be
investigated. Our study of transverse bifurcations is by far incomplete. We
have assumed that in a neighbourhood of the cycle the behaviour of trajectories
is accurately approximated by a collection of maps in which only
linearisations are taken into account. This is true in a neighbourhood, where
the linear part is significantly larger than the nonlinearity, but when
a (transverse) eigenvalue of the linearisation vanishes, the neighbourhood
of a steady state where this holds true shrinks to zero. For a more detailed
analysis of the bifurcation, some non-linear terms in the local expansion
of the r.h.s. $f$ should also be included into the local map.

In subsection \ref{exa} we have presented two general examples of type Z
homoclinic cycles. The question arises whether type Z homoclinic cycles
of other kinds exist; if they do exist, their classification is wanted.
(Heteroclinic cycles are apparently too diverse for any such classification
to be useful.)

\subsection*{Acknowledgements}

My research was financed in part by the grants ANR-07-BLAN-0235 OTARIE
from Agence Nationale de la Recherche, France, and 11-05-00167-a from
the Russian foundation for basic research.

\end{document}